\documentclass[epjCONF]{svjour}

\usepackage{graphicx}
\usepackage[varg]{txfonts} 
\usepackage[latin1]{inputenc}

\usepackage{bm}
\usepackage{savesym}
\savesymbol{iint}
\savesymbol{iiint}
\savesymbol{iiiint}
\savesymbol{idotsint}
\usepackage[centertags,leqno,sumlimits,intlimits,namelimits]{amsmath}
\usepackage{amsfonts,amsbsy,amsxtra}
\usepackage{amssymb,latexsym}
\restoresymbol{AMS}{iint}
\restoresymbol{AMS}{iiint}
\restoresymbol{AMS}{iiiint}
\restoresymbol{AMS}{idotsint}
\usepackage{color}
\usepackage{dcolumn}
\usepackage{textcomp}
\usepackage{booktabs}
\usepackage{soul}
\usepackage{ifthen}
\usepackage[
  unicode=true,
  pdftex=true,
  plainpages=false,
  bookmarks=true,
  bookmarksnumbered=true,
  bookmarksopen=true,
  bookmarksopenlevel=2,
  colorlinks=true,
  anchorcolor=gray50,
  linkcolor=black,
  urlcolor=black,
  citecolor=black,
  hyperindex=true
]{hyperref}
\usepackage{lineno}

\sethlcolor{green}
\setstcolor{red}

\newcommand{\figref}[1]{{Fig.~\ref{fig:#1}}}
\newcommand{\Figref}[1]{{Figure~\ref{fig:#1}}}

\newcommand{\tabref}[1]{{Table~\ref{tab:#1}}}
\newcommand{\Tabref}[1]{{Table~\ref{tab:#1}}}

\newcommand{\others}{\textit{et al.}}

\newcommand{\antibar}[1]{%
  \overline{#1}%
}

\newcommand{\Pg}{\ensuremath{g}}
\newcommand{\PReg}{\ensuremath{\mathbb{R}}}
\newcommand{\PPom}{\ensuremath{\mathbb{P}}}
\renewcommand{\Pr}{\ensuremath{\rho}}
\newcommand{\Prthree}{\ensuremath{\rho_3}}

\newcommand{\Paone}{\ensuremath{a_1}}
\newcommand{\Patwo}{\ensuremath{a_2}}
\newcommand{\Pafour}{\ensuremath{a_4}}
\newcommand{\Pbone}{\ensuremath{b_1}}
\newcommand{\Pfzero}{\ensuremath{f_0}}
\newcommand{\Pfone}{\ensuremath{f_1}}
\newcommand{\Pftwo}{\ensuremath{f_2}}

\newcommand{\Ppi}{\ensuremath{\pi}}
\newcommand{\Ppione}{\ensuremath{\pi_1}}
\newcommand{\Ppitwo}{\ensuremath{\pi_2}}
\newcommand{\Ppiz}{\ensuremath{\pi^0}}
\newcommand{\Ppip}{\ensuremath{\pi^+}}
\newcommand{\Ppim}{\ensuremath{\pi^-}}

\newcommand{\Peta}{\ensuremath{\eta}}
\newcommand{\PK}{\ensuremath{K}}

\newcommand{\PKs}{\ensuremath{K^0_S}}
\newcommand{\PKl}{\ensuremath{K^0_L}}
\newcommand{\PKp}{\ensuremath{K^+}}
\newcommand{\PKm}{\ensuremath{K^-}}

\newcommand{\Pqq}{\ensuremath{q}}
\newcommand{\Pqqbar}{\ensuremath{\antibar{q}}}
\newcommand{\Pqu}{\ensuremath{u}}
\newcommand{\Pqd}{\ensuremath{d}}
\newcommand{\Pqs}{\ensuremath{s}}
\newcommand{\Pp}{\ensuremath{p}}
\newcommand{\Ppbar}{\ensuremath{\antibar{p}}}
\newcommand{\Pn}{\ensuremath{n}}
\newcommand{\PN}{\ensuremath{N}}

\newcommand{\PPb}{\ensuremath{\text{Pb}}}
\newcommand{\twopion}{\ensuremath{\Ppip\Ppim}}
\newcommand{\threepion}{\ensuremath{\Ppim\Ppip\Ppim}}
\newcommand{\fourpion}{\ensuremath{\Ppip\Ppim\Ppip\Ppim}}
\newcommand{\fivepion}{\ensuremath{\Ppim\Ppip\Ppim\Ppip\Ppim}}

\newcommand{\lrBrk}[1]{{\left[{#1}\right]}}
\newcommand{\lrabs}[1]{{\left|{#1}\right|}}
\newcommand{\abs}[1]{{|{#1}|}}
\newcommand{\ket}[1]{{\vert{#1}\rangle}}

\newcommand{\xF}{\ensuremath{x_F}}

\newcommand{\measresult}[4]{%
  \ensuremath{#1%
    \ifthenelse{\equal{#2}{}}%
    {}%
    {\pm #2%
      \ifthenelse{\equal{#3}{}}%
      {}%
      {_\text{stat.}}%
    }%
    \ifthenelse{\equal{#3}{}}%
    {}%
    {\pm #3_\text{syst.}}\text{#4}%
  }%
}

\newcommand{\jpc}{\ensuremath{J^{PC}}}
\newcommand{\me}{\ensuremath{M^\varepsilon}}

\newcommand{\wavespec}[7]{\ensuremath{#1^{#2#3}}\!\!
  \ensuremath{#4^{#5}}\!\! \ensuremath{[#6] #7}} 

\newcommand{\gevc}{~\ensuremath{\text{GeV}\! / c}}
\newcommand{\gevcsq}{~\ensuremath{(\text{GeV}\! / c)^2}}
\newcommand{\mevcc}{~\ensuremath{\text{MeV}\! / c^2}}
\newcommand{\gevcc}{~\ensuremath{\text{GeV}\! / c^2}}
\newcommand{\tenpow}[2][]{%
  \ifthenelse{\equal{#1}{}}
  {\ensuremath{10^{#2}}}
  {\ensuremath{{#1} \cdot 10^{#2}}}
}

\session-title{%
19$^{\textnormal{\footnotesize th}}$ International %
IUPAP Conference on Few-Body Problems in Physics%
}

\begin{document}


\title{Precision Meson Spectroscopy at COMPASS}

\author{B. Grube\thanks{\email{bgrube@ph.tum.de}} for
  the COMPASS Collaboration}

\institute{Excellence Cluster Universe, Technische Universit\"at
  M\"unchen, Boltzmannstr. 2, 85748 Garching, Germany.}

\abstract{%
  We present first results of a partial wave analysis of the
  diffractive reaction $\Ppim\,\PPb \to \threepion\,\PPb$ based on
  data from the COMPASS experiment taken during a pilot run in 2004
  using a 190\gevc\ \Ppim\ beam on a lead target. The analysis was
  performed in the region of squared four-momentum transfer $t'$
  between 0.1 and 1.0\gevcsq. The \threepion\ final state shows a rich
  spectrum of well-known resonances. In addition a spin-exotic $\jpc =
  1^{-+}$ state with significant intensity was observed at 1.66\gevcc\
  in the $\Pr(770)\,\Ppi$ decay channel in natural parity exchange. The
  resonant nature of this state is manifest in the mass dependence of
  its phase difference to $\jpc = 1^{++}$ and $2^{-+}$ waves. The
  measured resonance parameters are consistent with the disputed
  $\Ppione(1600)$. An outlook on the analyses of the much larger data
  set taken during 2008 and 2009 is given.
}

\maketitle

\section{Introduction}

The na\"ive Constituent Quark Model (CQM) describes light mesons as
bound color-singlet states of quarks and antiquarks with flavors \Pqu,
\Pqd, and \Pqs\ grouped into $\text{SU}(3)_\text{flavor}$
multiplets. In the CQM the spin $J$, parity $P$, and charge
conjugation $C$ of a meson are given by
\begin{equation}
  J = \abs{L - S}, \ldots, L + S; \; P = (-1)^{L + 1} \;
  \text{and}\; C = (-1)^{L + S}
\end{equation}
where $L$ is the relative orbital angular momentum of quark and
antiquark and $S = 0, 1$ the total intrinsic spin of the
$\Pqq\Pqqbar'$ pair. In addition a meson is characterized by its
isospin $I$ and $G$-parity which is defined as
\begin{equation}
  G = (-1)^{I + L + S}
\end{equation}
Both quantum numbers are conserved in strong interactions.

Despite of its simplicity, the CQM works surprisingly well and
explains a good part of the observed light meson spectrum as well as
some of the meson properties, although it does not make any
assumptions about the nature of the binding force. Quantum
ChromoDynamics (QCD) describes the strong interaction between colored
quarks by the exchange of gluons, which are colored themselves. Even
though it is still not clear, how the confinement of quarks and gluons
into hadrons is generated by the QCD Lagrangian, its structure
suggests that there are color-sing\-let states beyond
$\ket{\Pqq\Pqqbar'}$. In particular one expects gluonic degrees of
freedom, that reflect the non-Abelian character of QCD, to manifest
themselves in the meson spectrum. Mesonic states beyond the CQM are
classified into hybrids, glueballs, and multi-quark
states~\cite{klempt_zaitsev}. Hybrids are $\ket{\Pqq\Pqqbar'\Pg}$
resonances with constituent glue that contributes to the quantum
numbers of the hadron. The $\Pqq\Pqqbar'$ pair is in a color-octet
configuration which is neutralized in color by the excited
gluon. Glueballs are pure gluonic bound states
$\ket{\Pg\Pg}$. Multi-quark states include tetraquarks and mesonic
molecules. Most of these states will be hardly distinguishable from
ordinary $\ket{\Pqq\Pqqbar'}$ states with the same \jpc\ and will mix
with them so that physically observable states are in general a linear
combination of $\ket{\Pqq\Pqqbar'}$ and additional basis states that
is experimentally very difficult to disentangle. An unambiguous
evidence for the existence of mesonic states beyond the CQM, as
allowed by QCD, would be the discovery of exotic states with quantum
numbers forbidden in the simple quark model. In the light-meson sector
so-called spin-exotic states with quantum numbers $\jpc = 0^{--}$,
$0^{+-}$, $1^{-+}$, $2^{+-}$, $3^{-+}$, $4^{+-}$, \ldots\ are a
particularly promising field of research.

In lattice QCD~\cite{glueball_lQCD} simulations the lightest glueball
is predicted to have ordinary scalar quantum numbers $\jpc = 0^{++}$
and a mass of about 1.7\gevcc. A possible experimental glueball
candidate is the $\Pfzero(1500)$ seen by the Crystal
Barrel~\cite{cb_glueball} and WA102~\cite{wa102_glueball}
experiments. The interpretation of the data is, however, complicated
by the mixing of the $\Pfzero(1500)$ with other states.

The lowest mass hybrid, on the other hand, is predic\-ted~\cite{hybrid}
to have exotic quantum numbers $\jpc = 1^{-+}$ so that it will not mix
with ordinary resonances. It is expected to have a mass between 1.3
and 2.2\gevcc and in the flux-tube model it preferentially decays into
$\Pbone(1235)\,\Ppi$ and \linebreak
$\Pfone(1285)\,\Ppi$~\cite{hybrid_fluxTube}. Three experimental
candidates for isovector spin-exotic $\jpc = 1^{-+}$ states were found
so far: The $\Ppione(1400)$ was seen in $\Ppim\PN \to \Ppim\Peta\PN$
by the E852~\cite{e852_pione1400} and VES~\cite{ves_pione1400}
experiments. Crystal Barrel observed a $1^{-+}$ \Peta\Ppi\ state in
$\Ppbar\Pn \to \Ppim\Ppiz\Peta$~\cite{cb_pione1400_1} and $\Ppbar\Pp
\to \Ppiz\Ppiz\Peta$~\cite{cb_pione1400_2} Dalitz plot
analyses. Another $1^{-+}$ state, the $\Ppione(1600)$, was seen at
higher mass by the E852 and VES experiments in \linebreak
\Pr\Ppi~\cite{e852_pione1600_rho,e852_pione1600_rho_1,ves_pione1600_rho},
$\Peta'\Ppi$~\cite{e852_pione1600_etaprime,ves_pione1600_etaprime},
$\Pfone\Ppi$~\cite{e852_pione1600_fone,ves_pione1600_5pi}, and
$\Pbone\Ppi$~\cite{e852_pione1600_omega,ves_pione1600_5pi} decay modes
in peripheral \Ppim\Pp\ interactions.  It was also reported in
$\Ppbar\Pp \to \Pbone\Ppi\Ppi$~\cite{baker}. The resonant nature of
both $1^{-+}$ states is still discussed controversially in the
community~\cite{ves_pione1400,ves_pione1600_5pi}. In particular the
observation of the $\Ppione(1600)$ in the \Pr\Ppi\ decay channel is
heavily disputed. A different analysis of a larger E852 data set
showed that an extension of the wave set, used to model the data,
makes the $\Ppione(1600)$ signal
disappear~\cite{e852_pione1600_rho_2}. A third $1^{-+}$ state, the
$\Ppione(2000)$, was seen only by the E852 experiment in the
\Pfone\Ppi~\cite{e852_pione1600_fone} and \linebreak
\Pbone\Ppi~\cite{e852_pione1600_omega} decay channel.

The COMPASS experiment sets out to settle these and other issues in
light-meson spectroscopy by studying meson production in diffractive
and central production reactions using high-energetic hadron beams on
various fixed targets.

\begin{figure}[!t]
  \centering
  \includegraphics[width=0.8\columnwidth]{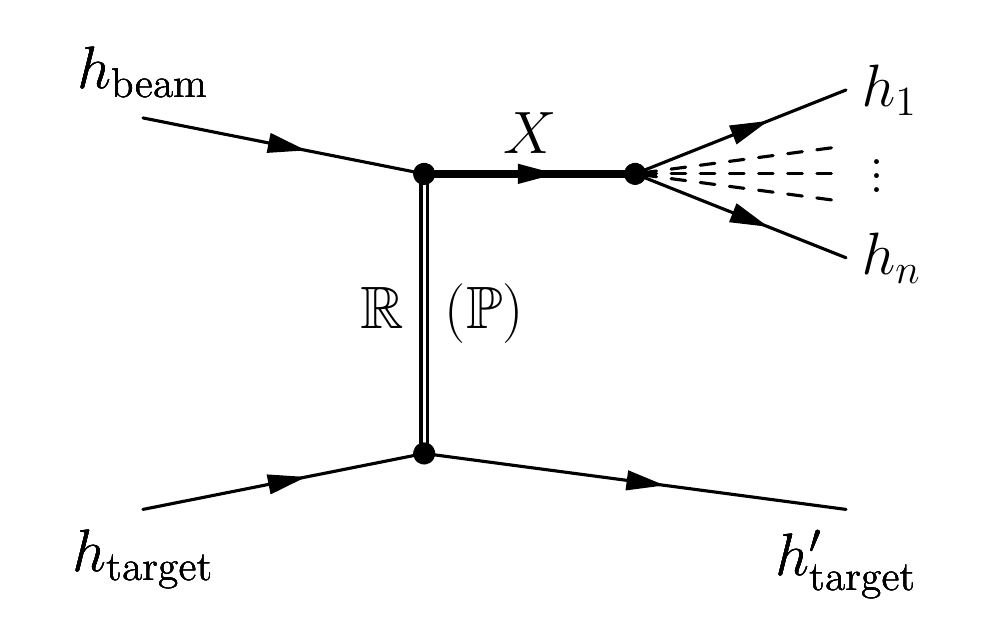}
  \caption{Production of resonance $X$ in diffractive scattering of the
    beam particle $h_\text{beam}$ off the target particle
    $h_\text{target}$ and its subsequent decay into $n$ hadrons $h_1,
    \ldots, h_n$. The interaction is mediated by exchange of a Reggeon
    \PReg\ and leaves target particle intact. At high energies
    the Pomeron \PPom\ is the dominant Regge-trajectory.}
  \label{fig:diffractive_production}
\end{figure}

Diffractive reactions are known to exhibit a rich spectrum of produced
states. In diffractive meson production the target particle remains
intact and interacts with the incident beam particle via $t$-channel
Reggeon exchange. In this process the beam particle $h_\text{beam}$ is
excited to some resonance $X$ which then dissociates into the final
state as illustrated in \figref{diffractive_production}:
\begin{equation}
  h_\text{beam} + h_\text{target} \to X + h_\text{target}'
  \quad \text{and} \quad X \to h_1 + \ldots + h_n
\end{equation}
The process can be characterized by two kinematic variables: the
square of the total center-of-mass energy, $s = (p_\text{beam} +
p_\text{target})^2$, and the squared four-momentum transfer from the
incoming beam particle to the outgoing system $X$, $t = (p_\text{beam}
- p_X)^2$. It is custom practice to use the variable $t' = \abs{t} -
\abs{t}_\text{min}$ instead of $t$, where $\abs{t}_\text{min}$ is the
minimum value of $\abs{t}$ allowed by kinematics for a given $X$
invariant mass $m_X$. The value of $\abs{t}_\text{min}$ is small but
larger than zero, because of the additional longitudinal four-momentum
transfer which is required by $m_X > m_\text{beam}$. At high beam
energies the diffraction is dominated by Pomeron
exchange~\cite{pomeron} so that isospin and $G$-parity of the
intermediate state $X$ are the same as that of the beam particle. In
addition the final state particles are produced mostly at small angles
with respect to the beam direction which requires a high angular
resolution and coverage of the detector.

\begin{figure}[!t]
  \centering
  \includegraphics[width=0.8\columnwidth]{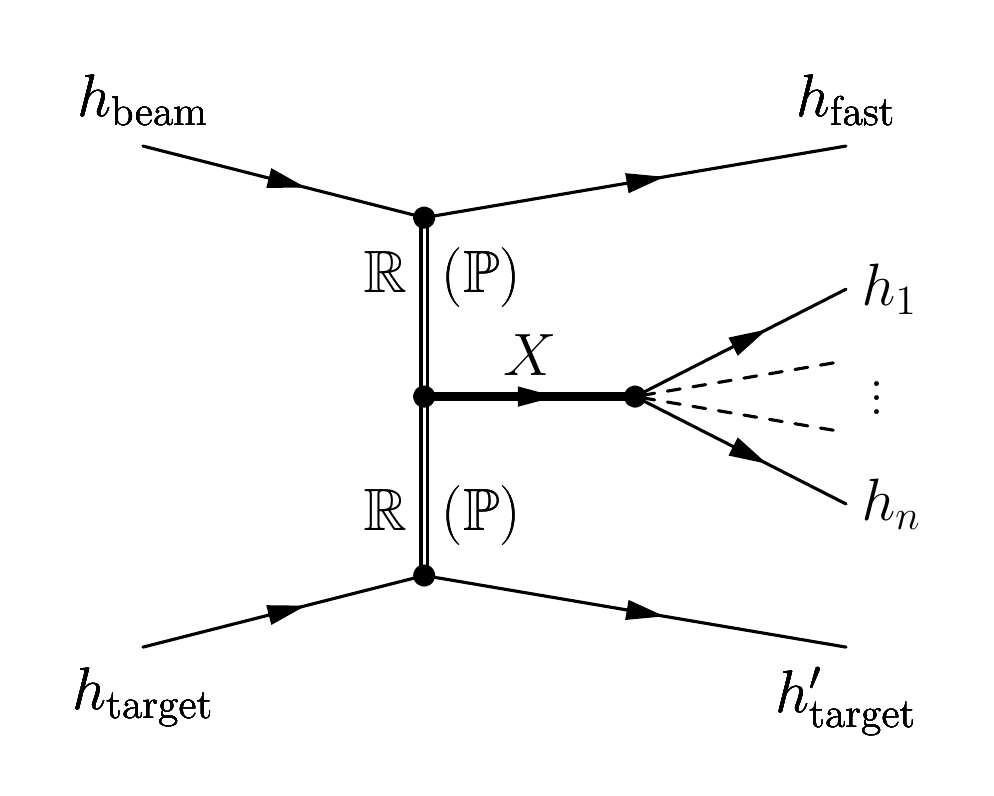}
  \caption{Central production of meson $X$ in Reggeon-Reggeon fusion
    and its subsequent decay into $n$ hadrons $h_1, \ldots, h_n$. Both
    beam and target particles stay intact.}
  \label{fig:central_production}
\end{figure}

In central production reactions the mesonic state is produced in
Reggeon-Reggeon fusion, where both beam and target particle remain
unaltered. This is illustrated in \figref{central_production}. These
reactions will enhance scalar mesons and in
particular to the $\Pfzero(1500)$. It is also believed that in central
production gluonic degrees of freedom are enriched which makes it a
promising reaction for glueball searches. The produced resonance $X$
carries only a small fraction $\xF \approx 0$ of the maximum possible
longitudinal momentum $p_L^\text{max}$ in the center-of-mass frame,
where $\xF = p_L / p_L^\text{max} \approx 2 p_L / \sqrt{s}$. The
scattered beam hadron has $\xF \approx 1$ and the target $\xF \approx
-1$. In fixed target geometry this means that the scattered beam
particle appears as the leading hadron $h_\text{fast}$ in the event,
while the target recoil proton is slow. The centrally produced system is
separated from the leading hadron and the recoil particle by rapidity
gaps.

\section{Experimental Setup}

Finding mesonic states that do not fit into the CQM in the light-quark
sector is an experimentally challenging task. There is a high density
of broad overlapping states so that the signal can only be extracted
from interference effects. This requires large data sets and a
complete phase space coverage of the experiment. The COmmon Muon and
Proton Apparatus for Structure and Spectroscopy (COM- \linebreak
PASS)~\cite{compass} is a fixed-target experiment at the CERN Super
Proton Synchrotron (SPS) and is well suited for this task: It is a
two-stage spectrometer built to measure at high beam intensities. The
detector covers a wide range of scattering angles and particle momenta
and has a high angular resolution. In the target region scintillating
fibers and planes of silicon microstrip detectors are used for beam
definition and vertexing. The setup has two large dipole magnets with
1.0 and 5.5~Tm bending power which both are surrounded by staggered
tracking detectors with increasing granularity and resolution towards
the beam axis. The two spectrometer stages are equipped with hadronic
and electromagnetic calorimeters. This enables COMPASS to reconstruct
final states with charged as well as neutral particles like \Ppiz,
\Peta, and $\Peta'$. In addition the first stage features a Ring
Imaging CHerenkov (RICH) detector able to separate charged final state
pions and kaons in the momentum range between 5 and 50\gevc.

The M2 beamline that feeds the COMPASS experiment is very versatile
and can deliver high-intensity secondary hadron as well as polarized
tertiary muon beams which are produced at a production target by the
incoming 400\gevc\ proton beam from the SPS. The hadron beams can have
a momentum of up to 300\gevc\ and a maximum intensity of
\tenpow[4]{7}~$\text{sec}^{-1}$. The positive hadron beam consists of
71.5~\% \Pp, 25.5~\% \Ppip, and 3.0~\% \PKp, the negative beam has
96.0~\% \Ppim and 3.5~\% \PKm.

The suitability of the COMPASS setup for meson spectroscopy was
studied in a short pilot run in 2004, where a 190\gevc\ \Ppim\ beam
was shot onto a fixed lead target. Based on the experience from this
run the spectrometer was upgraded in order to address the challenges
of the hadron spectroscopy physics program. A Recoil Proton Detector
(RPD) was installed around a 40~cm long liquid hydrogen target. The
RPD measures the time-of-flight of the recoil protons using two
barrels of scintillator slats and provides information for the trigger
decision. The readout electronics of the electromagnetic calorimeters
was upgraded and the central part of the second calorimeter was
equipped with 800 radiation-hard Shashlik blocks. In addition the
tracking close to the beam axis and the vertex definition were
improved by adding high-resolution PixelGEM detectors in the
spectrometer and cryogenic silicon microstrip detectors in the target
region, respectively. At the same time the material budget in the beam
region was reduced. Also the particle identification capabilities were
enhanced by utilizing the RICH detector in the first spectrometer
stage and two ChErenkov Differential counters with Achromatic Ring
focus (CEDAR) upstream of the target which are able to identify the
incoming beam particles.

With this enhanced setup COMPASS took diffractive and central
production data in 2008 and 2009 using \linebreak 190\gevc\ negative
and positive hadron beams on liquid hydrogen, nickel, tungsten, and
lead targets.  The goal of collecting about ten times the available
world statistics for both production reactions has been achieved.

\mathversion{bold}
\section{Diffractive Production of \threepion}
\mathversion{normal}

During the 2004 pilot run COMPASS was able to record within a few days
a data sample of the diffractive reaction
\begin{equation}
  \Ppim + \PPb \to \threepion + \PPb
\end{equation}
with competitive statistics using a 190\gevc\ \Ppim\ beam on a total
3~mm thick lead target. The \threepion\ final state was chosen for a
first analysis~\cite{quirin_meson08,quirin_phd,pione1600_paper},
because the controversial spin-exotic $\Ppione(1600)$ has been
observed in this channel.

\subsection{Data Sample and Event Selection}

The trigger for diffractive events selected one incoming and at least
two outgoing charged particles. In 2004 it consisted of several
components. Incoming beam particles were selected by a coincidence of
two scintillator discs with 5~cm diameter that were positioned
upstream of the target and centered about the beam axis. A veto system
with a 4~cm central hole rejected beam particles that were not
crossing the target material. Additional lead-scintillator ve\-to
counters placed downstream of the target suppressed events with
outgoing particles that were emitted under large angles so that they
would fall out of the acceptance. A system of three scintillation
detectors in the spectrometer, so-called beam killers, vetoed
non-interacting beam particles. In order to enrich diffractive events,
a scintillator disk with 5~cm diameter was placed after the target and
was used as a multiplicity counter, selecting events with at least two
forward charged particles. In addition at least one cluster with
minimum energy deposit of 6~GeV was required in the hadronic
calorimeter in the second spectrometer stage.

\begin{figure}[!b]
  \centering
  \includegraphics[width=\columnwidth]{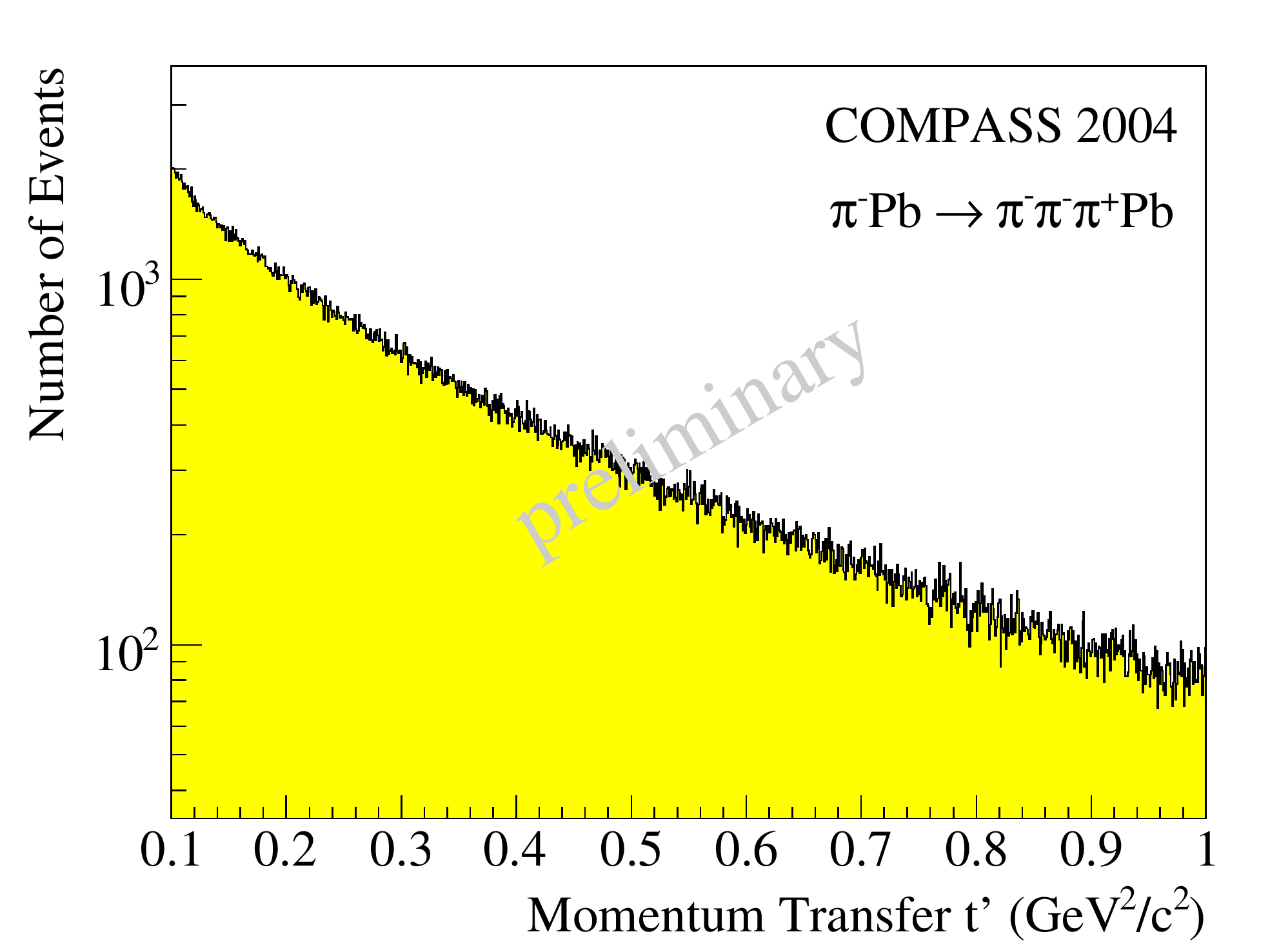}
  \caption{Spectrum of the squared four-momentum transfer $t' =
    \abs{t} - \abs{t}_\text{min}$ in logarithmic scale. The data
    exhibit a roughly exponential behavior.}
  \label{fig:tPrime_high}
\end{figure}

\begin{figure}[!t]
  \centering
  \includegraphics[width=\columnwidth]{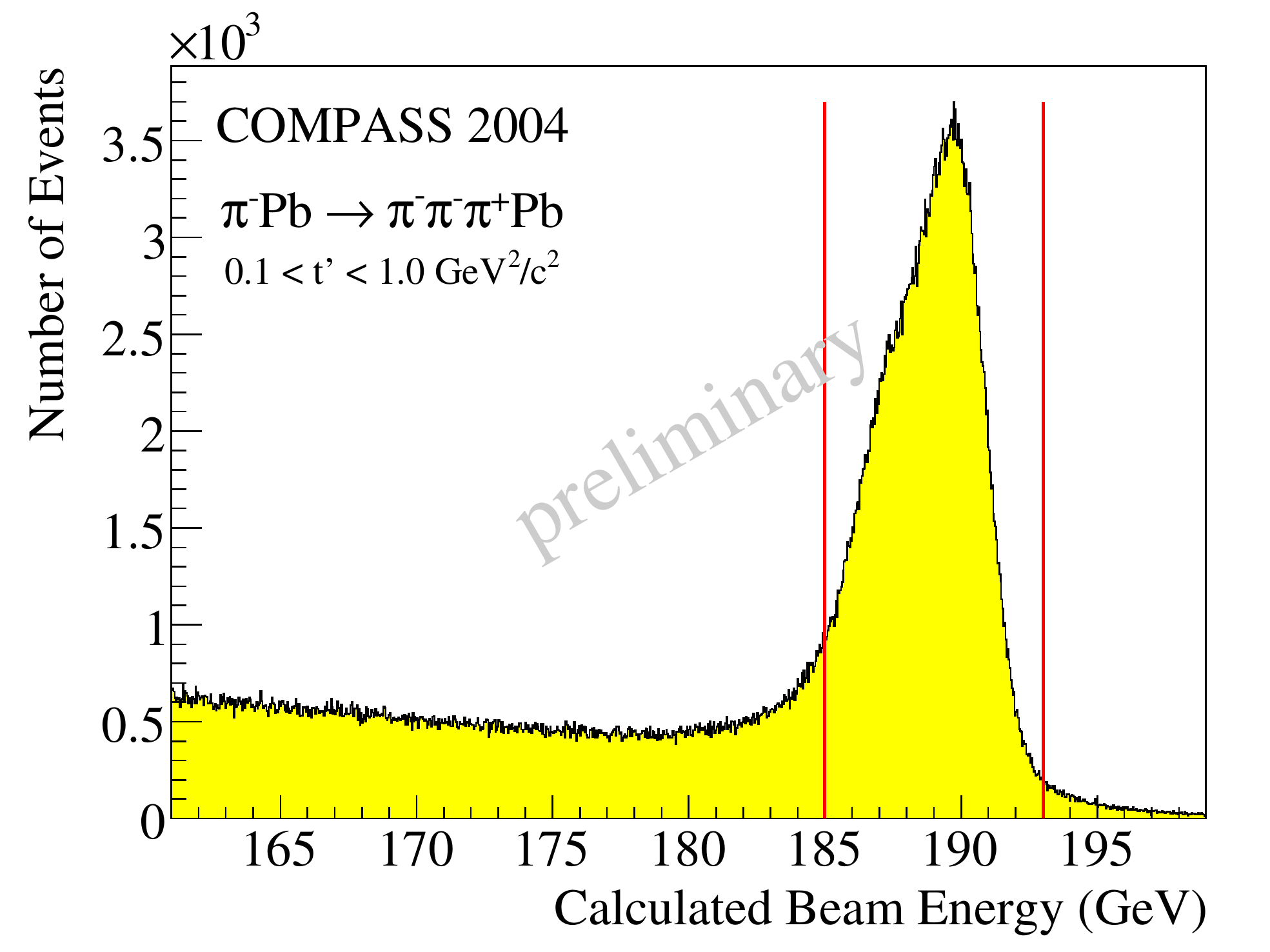}
  \caption{Distribution of the beam energy $E_\text{beam}$ calculated
    from $m_X$, $E_X$, and $\theta$ for the $t'$ range between 0.1 and
    1.0\gevcsq. The exclusivity cut is indicated by the vertical
    lines.}
  \label{fig:exclusivity}
\end{figure}

The 2004 data sample contains about \tenpow[87.7]{6} events from the
diffractive trigger which were subject to further offline event
selection. Events were required to have exactly one primary vertex
with one incoming beam and three outgoing charged tracks. The primary
vertex position was constraint to the target region and the three
outgoing tracks were required to have a charge sum of
$-1$. Diffractive events were enriched by an exclusivity cut. However,
the spectrometer setup was not able to measure the energy of the
incident beam particle event-by-event, only the direction of the beam
particle and thus the scattering angle $\theta$ was determined
precisely by a silicon microstrip beam telescope. In the 2004 setup
also the target recoil particle was not detected. For small momentum
transfer, large beam energy, and assuming specific target and recoil
masses the beam energy $E_\text{beam}$ and the momentum transfer $t'$
can be calculated in good approximation from the well-measured
quantities $m_X$, $E_X$, and $\theta$. For the analysis the $t'$
region between 0.1 and 1.0\gevcsq\ was selected, where the
$\Ppione(1600)$ was reported in the past. In this $t'$ region the beam
particles were assumed to scatter off quasi-free nucleons inside the
\PPb~nuclei of the target. The validity of this assumption is
supported by the roughly exponential falling $t'$-spectrum shown in
\figref{tPrime_high} which has a slope parameter compatible with the
nucleon radius. Exclusive \threepion\ production was selected by
requiring the reconstructed beam energy $E_\text{beam}$ to be maximum
4~GeV off the mean beam energy of 189~GeV (see
\figref{exclusivity}). About 420\,000 events in the mass range between
0.5 and 2.5\gevcc\ pass all of the above selection cuts.

In the 2004 pilot run there was no particle identification for the
incoming beam particles. The non-pionic component of the beam of about
5~\% consists mainly of \PKm. Their decays and diffractive reactions
constitute part of the background in the analysis. Also for the final
state no particle identification was applied.

\subsection{Partial-Wave Analysis}

\begin{figure}[!t]
  \centering
  \includegraphics[width=\columnwidth]{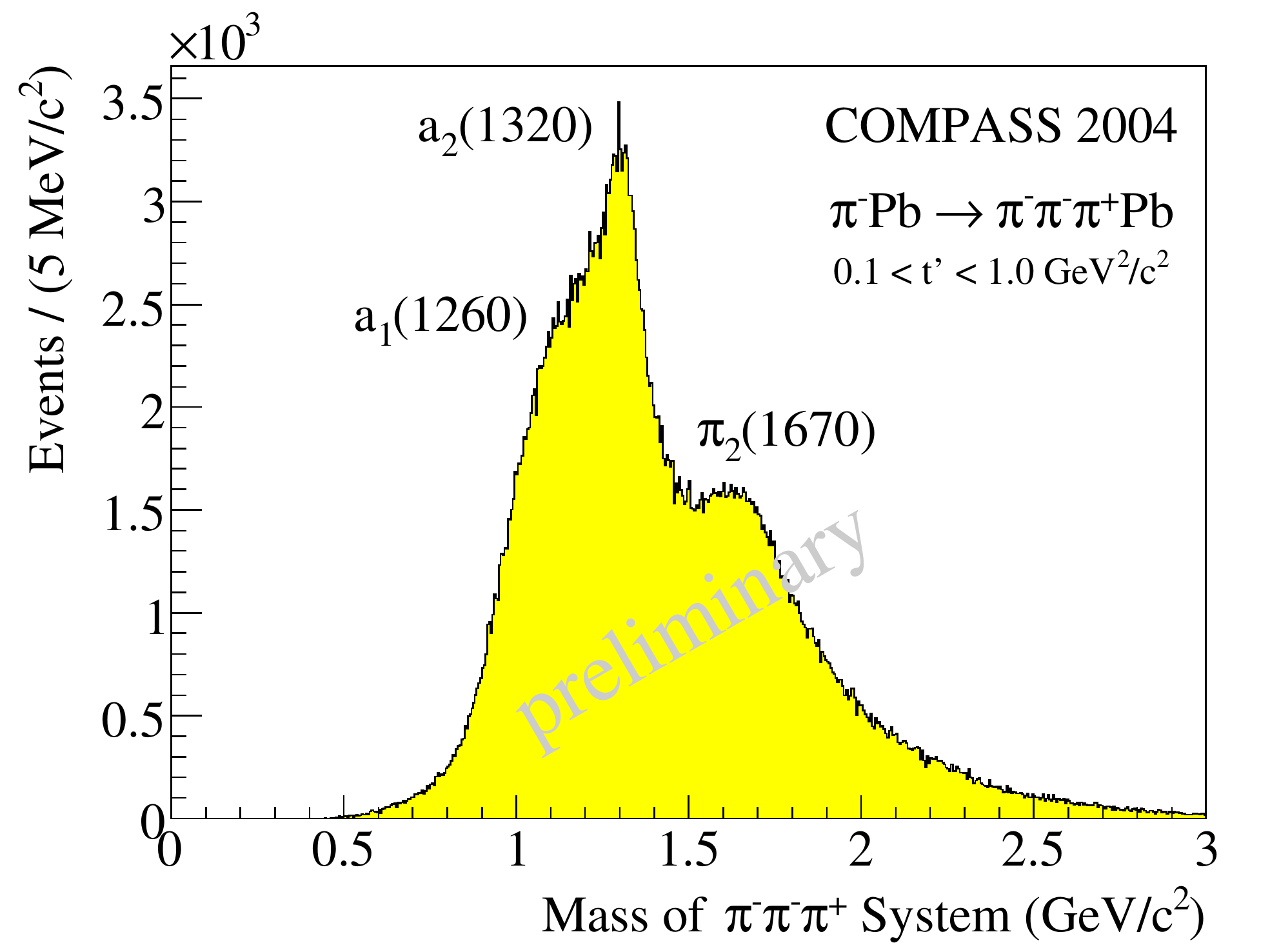}
  \caption{\threepion\ invariant mass distribution of the selected
    data sample for $t' \in [0.1, 1.0]\gevcsq$.}
  \label{fig:threePiMass}
\end{figure}

\Figref{threePiMass} shows the \threepion\ invariant mass distribution
of the selected data sample. It exhibits clear structures in the mass
region of the well-known resonances $\Paone(1260)$, \linebreak
$\Patwo(1320)$, and $\Ppitwo(1670)$. In order to find and disentangle
the various resonances in the data, a partial-wave analysis (PWA) was
performed.

\begin{figure}[!t]
  \centering
  \includegraphics[width=0.8\columnwidth]{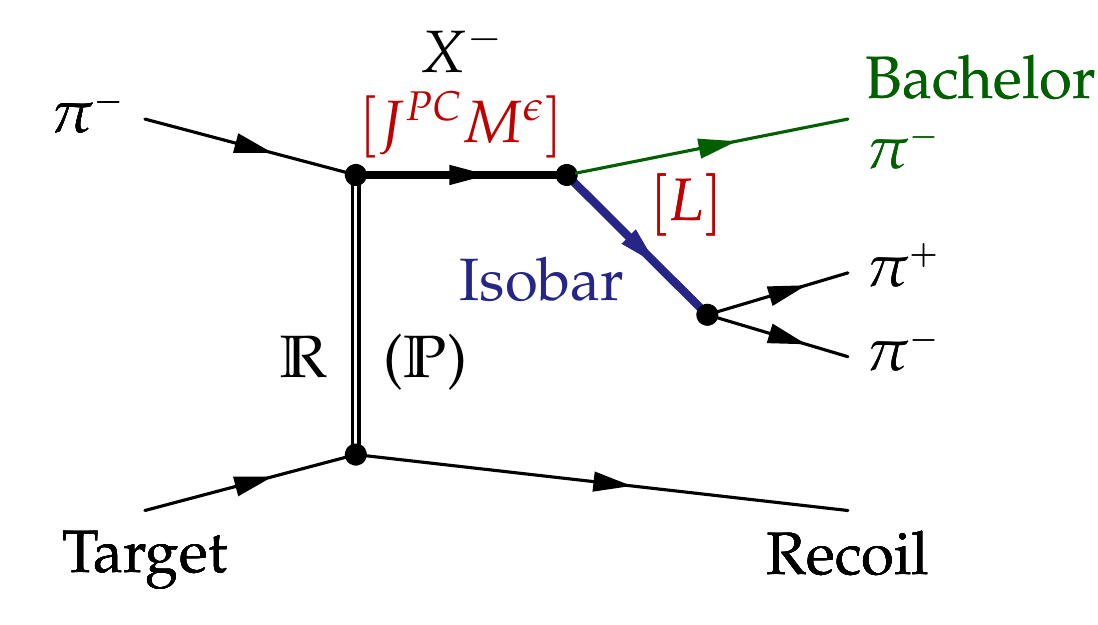}
  \caption{A partial wave in the isobar model. The diffractively
    produced $X^-$ with quantum numbers $\jpc M^\epsilon$ decays first
    into an isobar and a bachelor \Ppim. The isobar has spin $S$ and a
    relative orbital angular momentum $L$ with respect to the
    $\Ppim_\text{bachelor}$. The isobar then decays in a second step
    into \twopion.}
  \label{fig:isobar}
\end{figure}

The PWA employed in this analysis is based on two basic assumptions.
We assume that the total cross section factorizes into a resonance and
a recoil vertex and we use the isobar model to describe the $X^-$
decay. This by definition neglects any final state interaction of the
outgoing pions or the isobar with the target as well as within the
\threepion\ system. The isobar model decomposes the decay $X^- \to
\threepion$ into a chain of successive two-body decays: First the
$X^-$ with quantum numbers \jpc\ decays into a di-pion resonance, the
so-called isobar, and a bachelor pion $\Ppim_\text{bachelor}$. The
isobar has spin $S$ and a relative orbital angular momentum $L$ with
respect to $\Ppim_\text{bachelor}$. $L$ and $S$ couple to the $X^-$
spin $J$. Second the isobar decays into \twopion. This is illustrated
in \figref{isobar}. In general the \threepion\ system has isospin $I
\geq 1$. Since there are no known flavor-exotic light-quark mesons and
Pomeron exchange is assumed, we set $I = 1$. A state with odd number
of pions has negative $G$-parity so that the value of the charge
conjugation $C$ is fixed by $G = C (-1)^I$ to be positive.

The spin-density matrix $\rho_{i, i'}^\epsilon$ is parameterized in
terms of production amplitudes $V_{i r}^\epsilon$:
\begin{equation}
  \rho_{i, i'}^\epsilon = \sum_r^{N_r} V_{i r}^\epsilon V_{i' r}^{\epsilon *}
\end{equation}
where $i = \jpc M^\epsilon[\text{decay}]L$ and $\epsilon$ denote a
particular partial-wave amplitude that is characterized by the isobar
state, the decay orbital angular momentum $L$, the spin quantum
numbers \jpc\ of the $X^-$ state, and by the spin projection
$M^\epsilon$ of $J$. The amplitudes are constructed in the
reflectivity basis~\cite{reflectivity}, where $M \geq 0$ and where the
reflectivity $\epsilon = \pm 1$ describes the symmetry under
reflection through the production plane. This basis is convenient,
because, due to $P$ parity conservation, amplitudes with different
reflectivities do not interfere. Furthermore at high center-of-mass
energies $\sqrt{s}$ the reflectivity corresponds to the naturality of
the exchanged Reggeon. The rank $N_r$ of the spin density matrix is
set to two to account for the helicity flip and non-flip amplitudes at
the baryon vertex assuming that the target nucleon stays intact.

The observed intensity is parameterized as a coherent and incoherent
sum over the partial-wave amplitudes~\cite{reflectivity}:
\begin{equation}
  I(\tau; m_X) = \eta(\tau) \sum_{\epsilon = \pm 1} \sum_r^{N_r}
  \lrabs{\sum_i^\text{waves}\!\! V_{i r}^\epsilon\, \psi_i^\epsilon(\tau; m_X)}^2
\end{equation}
Here $\tau$ represents the five phase space coordinates that
completely describe the three-body kinematics. They are measured for
each event and are used to calculate the decay amplitudes
$\psi_i^\epsilon$ which are constructed using non-relativistic Zemach
tensors~\cite{zemach_1,zemach_2} and do not contain any free
parameters. The decay of the $X^-$ is described in its rest system
using the Gottfried-Jackson frame~\cite{GJ_frame} with the $z$-axis
along the beam particle direction and the $y$-axis perpendicular to
the production plane spanned by the beam and the recoil particle. The
angular distribution of the isobar decay is defined in the canonical
system obtained by pure Lorentz boost to the isobars rest frame. The
production amplitudes $V_{i r}^\epsilon$ are complex numbers that are
determined from extended maximum likelihood fits to the data performed
in 40\mevcc\ wide bins in the three-pion invariant mass $m_X$ using a
program that was originally developed at Illinois~\cite{pwa_prog} and
later developed at Protvino and Munich. This so-called
``mass-independent'' fit takes into account the overall acceptance
$\eta$ as a function of the phase space variables. It does not include
any parameterization of the produced resonances $X^-$ and assumes
constant production strengths of the waves within the $m_X$ bin. The
$t'$ dependence of the wave intensities was implemented by multiplying
the amplitudes with different $t'$-dependent functions of the form
$f_i^\epsilon(t') \propto e^{-bt'}$ for $M = 0$ and $f_i^\epsilon(t')
\propto t'e^{-bt'}$ for $M = 1$, where the slope parameter $b$ was
determined from the data by fitting $t'$ slices.

\begin{figure}[!h]
  \centering
  \includegraphics[width=\columnwidth]{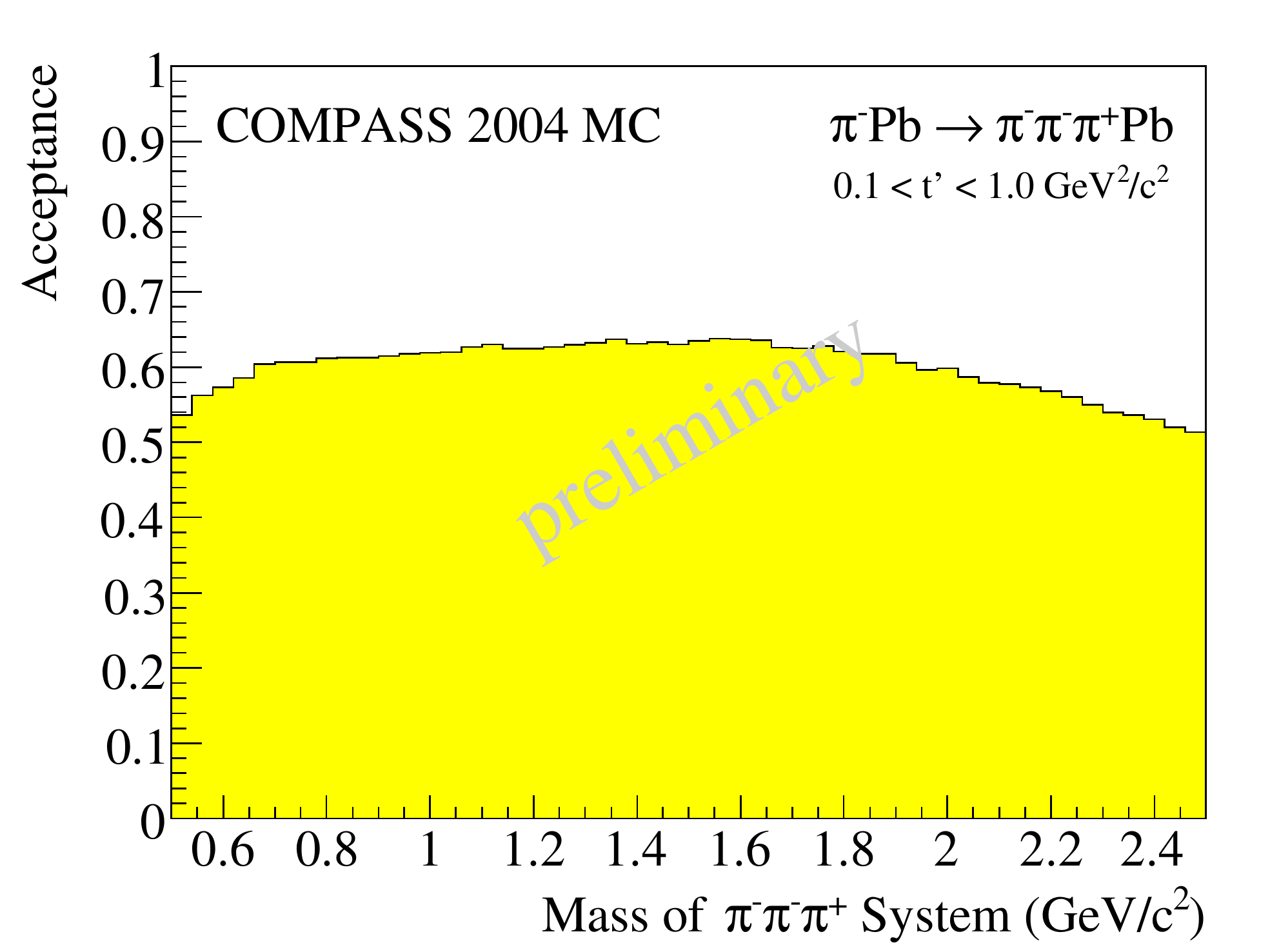}
  \caption{Overall acceptance as a function of the \threepion\
    invariant mass $m_X$.}
  \label{fig:acc_3piMass}
\end{figure}

\begin{figure}[!h]
  \centering
  \includegraphics[width=\columnwidth]{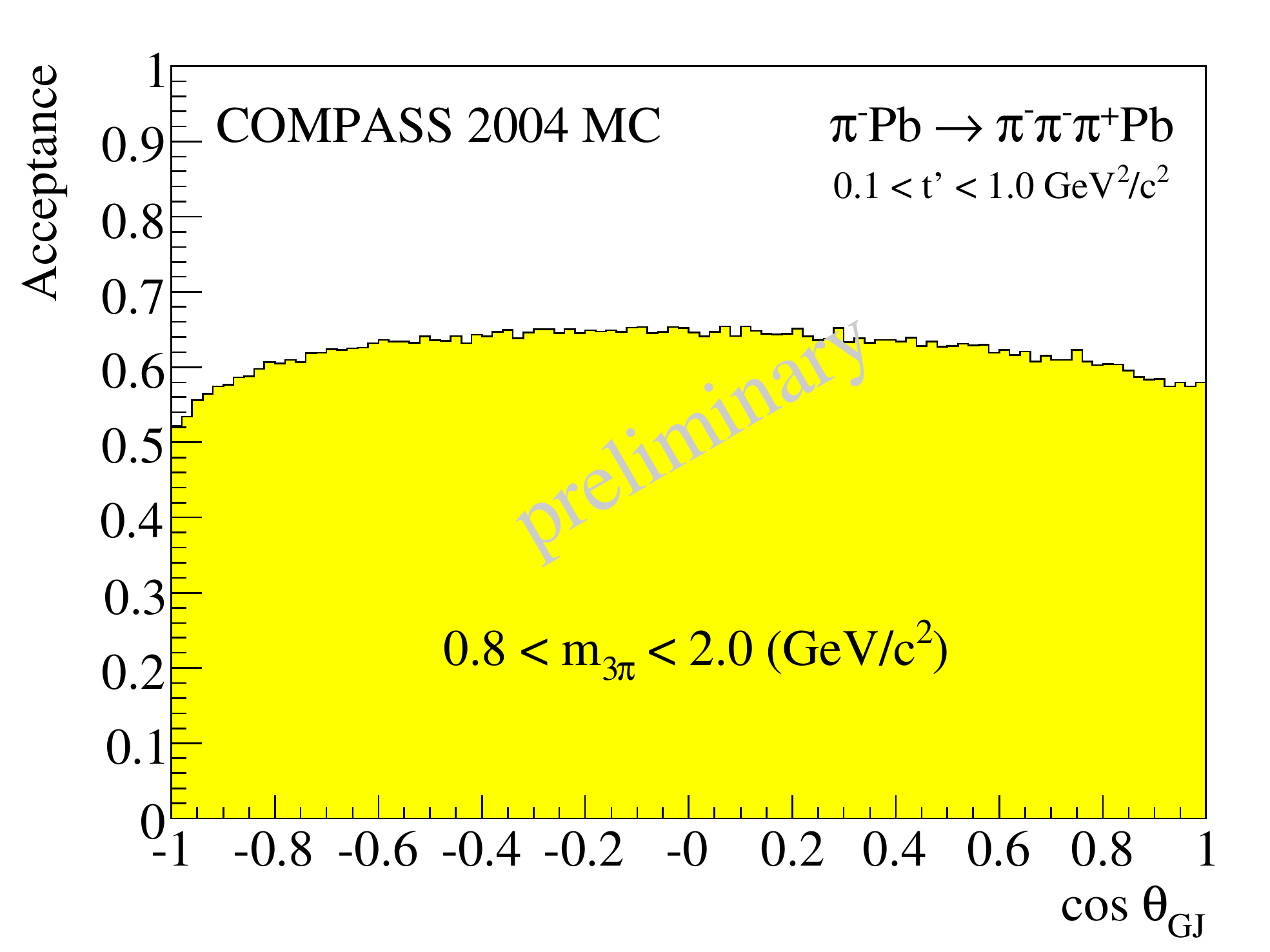}
  \caption{Overall acceptance as a function of the cosine of the polar
    angle $\theta_{GJ}$ of the isobar in the Gottfried-Jackson rest
    frame of the $X^-$.}
  \label{fig:acc_cosThetaGJ}
\end{figure}

The acceptance was estimated using a Monte Carlo simulation. Events
were generated uniformly according to three-body phase space with beam
kinematics and $t'$ distribution taken from the real data and then
processed through the detailed detector simulation and the full
reconstruction and event selection chain. \Figref{acc_3piMass} shows
that COMPASS has an excellent acceptance of more than 50~\% for
\threepion\ final states even in the region $m_X > 2\gevcc$. Moreover
the acceptance shows only weak dependence on the polar angle of the
isobar in the Gottfried-Jackson frame
(cf. \figref{acc_cosThetaGJ}). This is important for the PWA, since
in~\cite{e852_pione1600_rho_2} it was shown that a drop of the
acceptance towards $\cos \theta_{GJ} = \pm 1$ in combinations with
detector resolution effects may lead to significant leakage of
intensity between waves.

The PWA model includes five \twopion\
isobars~\cite{e852_pione1600_rho_1}: The amplitudes of the $\Pr(770)$,
$\Pftwo(1270)$, and $\Prthree(1690)$ where described using
relativistic Breit-Wigner line shape functions including
Blatt-Weisskopf barrier penetration fac- \linebreak
tors~\cite{blatt_weisskopf}. The \twopion\ $S$-wave is in general
dominated by a broad $\Pfzero(600)$ meson, a narrow $\Pfzero(980)$,
and several resonances above 1\gevcc. We use the parameterization
\linebreak from~\cite{ves_sigma} which is based on the ``M'' solution
from Au, Morgan, and Pennington~\cite{amp_sigma}, but with the
$\Pfzero(980)$ subtracted from the elastic \Ppi\Ppi\ amplitude and
added as a separate Breit-Wigner resonance.

Due to the angular momentum carried by the Pomeron that is exchanged
in diffractive reactions, the incoming \Ppim\ ($J^P = 0^{-}$) can be
excited to a state $X^-$ with different $J^P$, which is limited only
by the conservation laws of strong interaction. In the construction of
the PWA model waves with $J \leq 4$ and $M \leq 1$ were
considered. The final wave set of the model consists of 41~partial
waves plus one incoherent isotropic background wave, the so-called
``flat'' wave. \Tabref{waveset} lists all waves. Mostly positive
reflectivity waves, corresponding to natural parity exchange
production, are needed to describe the data. The wave set in
particular includes the \wavespec{2}{-}{+}{0}{+}{\Pr\Ppi}{F},
\wavespec{2}{-}{+}{1}{+}{\Pr\Ppi}{F}, and
\wavespec{2}{-}{+}{1}{+}{\Pr\Ppi}{P} waves that
in~\cite{e852_pione1600_rho_2} were found to be crucial for describing
the data and in addition made the exotic $\Ppione(1600)$ signal
disappear. Some of the waves have $m_X$ thresholds below which their
intensity was fixed to zero.

\begin{table}[!b]
  \caption{Wave set used for the mass-independent fit. The set consists
    of mostly positive reflectivity waves that correspond to natural
    parity exchange. The six waves printed in boldface are used in the
    mass-dependent fit.}
  \label{tab:waveset}
  \begin{tabular}{ccccc}
    \toprule
    $\bm{\jpc}$ &
    $\bm{\me}$ &
    $\bm{L}$ &
    \textbf{Isobar} $\bm{\pi}$ &
    \textbf{Threshold [$\bm{\text{GeV}\! /\! c^2}$]} \\
    \midrule[0.08em]
    $0^{-+}$      & $0^+$      & $S$      & $(\Ppi\Ppi)_\text{S} \Ppi$ &   --- \\
    $\bm{0^{-+}}$ & $\bm{0^+}$ & $\bm{S}$ & $\bm{\Pfzero \Ppi}$       & \textbf{1.400} \\        
    $0^{-+}$      & $0^+$      & $P$      & $\Pr \Ppi$                &   --- \\
    \midrule
    $\bm{1^{-+}}$ & $\bm{1^+}$ & $\bm{P}$ & $\bm{\Pr \Ppi}$ & \textbf{---} \\
    \midrule
    $\bm{1^{++}}$ & $\bm{0^+}$ & $\bm{S}$ & $\bm{\Pr \Ppi}$           & \textbf{---} \\
    $1^{++}$      & $0^+$      & $P$      & $\Pftwo \Ppi$             & 1.200 \\
    $1^{++}$      & $0^+$      & $P$      & $(\Ppi\Ppi)_\text{S} \Ppi$ & 0.840 \\
    $1^{++}$      & $0^+$      & $D$      & $\Pr \Ppi$                & 1.300 \\
    $1^{++}$      & $1^+$      & $S$      & $\Pr \Ppi$                &   --- \\
    $1^{++}$      & $1^+$      & $P$      & $\Pftwo \Ppi$             & 1.400 \\
    $1^{++}$      & $1^+$      & $P$      & $(\Ppi\Ppi)_\text{S} \Ppi$ & 1.400 \\
    $1^{++}$      & $1^+$      & $D$      & $\Pr \Ppi$                & 1.400 \\
    \midrule
    $\bm{2^{-+}}$ & $\bm{0^+}$ & $\bm{S}$ & $\bm{\Pftwo \Ppi}$        & \textbf{1.200} \\
    $2^{-+}$      & $0^+$      & $P$      & $\Pr \Ppi$                & 0.800 \\
    $2^{-+}$      & $0^+$      & $D$      & $\Pftwo \Ppi$             & 1.500 \\
    $2^{-+}$      & $0^+$      & $D$      & $(\Ppi\Ppi)_\text{S} \Ppi$ & 0.800 \\
    $2^{-+}$      & $0^+$      & $F$      & $\Pr \Ppi$                & 1.200 \\
    $2^{-+}$      & $1^+$      & $S$      & $\Pftwo \Ppi$             & 1.200 \\
    $2^{-+}$      & $1^+$      & $P$      & $\Pr \Ppi$                & 0.800 \\
    $2^{-+}$      & $1^+$      & $D$      & $\Pftwo \Ppi$             & 1.500 \\
    $2^{-+}$      & $1^+$      & $D$      & $(\Ppi\Ppi)_\text{S} \Ppi$ & 1.200 \\
    $2^{-+}$      & $1^+$      & $F$      & $\Pr \Ppi$                & 1.200 \\
    \midrule
    $2^{++}$      & $1^+$      & $P$      & $\Pftwo \Ppi$   & 1.500 \\
    $\bm{2^{++}}$ & $\bm{1^+}$ & $\bm{D}$ & $\bm{\Pr \Ppi}$ & \textbf{---} \\
    \midrule
    $3^{++}$ & $0^+$ & $S$ & $\Pr_3 \Ppi$  & 1.500 \\
    $3^{++}$ & $0^+$ & $P$ & $\Pftwo \Ppi$ & 1.200 \\
    $3^{++}$ & $0^+$ & $D$ & $\Pr \Ppi$    & 1.500 \\
    $3^{++}$ & $1^+$ & $S$ & $\Pr_3 \Ppi$  & 1.500 \\
    $3^{++}$ & $1^+$ & $P$ & $\Pftwo \Ppi$ & 1.200 \\
    $3^{++}$ & $1^+$ & $D$ & $\Pr \Ppi$    & 1.500 \\
    \midrule
    $4^{-+}$ & $0^+$ & $F$ & $\Pr \Ppi$ & 1.200 \\
    $4^{-+}$ & $1^+$ & $F$ & $\Pr \Ppi$ & 1.200 \\
    \midrule
    $4^{++}$      & $1^+$      & $F$      & $\Pftwo \Ppi$   & 1.600 \\
    $\bm{4^{++}}$ & $\bm{1^+}$ & $\bm{G}$ & $\bm{\Pr \Ppi}$ & \textbf{1.640} \\
    \midrule
    $1^{-+}$ & $0^-$ & $P$ & $\Pr \Ppi$    &   --- \\
    $1^{-+}$ & $1^-$ & $P$ & $\Pr \Ppi$    &   --- \\          
    $1^{++}$ & $1^-$ & $S$ & $\Pr \Ppi$    &   --- \\
    $2^{-+}$ & $1^-$ & $S$ & $\Pftwo \Ppi$ & 1.200 \\
    $2^{++}$ & $0^-$ & $P$ & $\Pftwo \Ppi$ & 1.300 \\
    $2^{++}$ & $0^-$ & $D$ & $\Pr \Ppi$    &   --- \\
    $2^{++}$ & $1^-$ & $P$ & $\Pftwo \Ppi$ & 1.300 \\
    \midrule
    Flat & --- & --- & --- & --- \\
    \bottomrule
  \end{tabular}
\end{table}

To ensure that the mass-independent fit found the global maximum of
the likelihood, up to 100 independent fits were performed for each
mass bin using random start parameters. Out of these fits the one with
the largest likelihood was selected. In case two or more fits have
ambiguous solutions that differ by less than one unit of likelihood,
the error of this bin was increased by the maximum difference of the
solutions, which in the plots is indicated by thick green bars.

After performing the mass-independent fit the mass dependence of the
production amplitudes of a subset of six waves (printed in boldface in
\tabref{waveset}) was fit in the range from 0.80 to 2.32\gevcc\ to a
model parameterized in terms of Breit-Wigner amplitudes:
\begin{equation}
  \rho_{ij}^\epsilon(m_X) = \sum_r^{N_r}\lrBrk{\sum_k^\text{resonances}\!\!\!\!\!
    c_{ikr}^\epsilon\, \text{BW}\!_k(m_X)}\, \lrBrk{\sum_l^\text{resonances}\!\!\!\!\!
    c_{jlr}^\epsilon\, \text{BW}\!_l(m_X)}^*
\end{equation}
Here $\text{BW}\!_i(m_X)$ is the relativistic Breit-Wigner amplitude
for resonance $i$:
\begin{equation}
  \text{BW}\!_i(m_X; M_{0}, \Gamma\!_{0}) = \frac{1}{m_X^2 - M_{0}^2 +
    i \Gamma\!_\text{tot}(m_X) M_{0}}
\end{equation}
where $\Gamma\!_\text{tot}(m_X) = \sum_n^{N_\text{decay}} \Gamma\!_n(m_X)$
is the total mass-dependent width of the resonance including phase
space factors and Blatt-Weisskopf barrier penetration factors
$B_{Ln}(q_n)$~\cite{blatt_weisskopf} for all $N_\text{decay}$ decay
channels:
\begin{equation}
  \Gamma\!_\text{tot}(m_X) = \sum_n^{N_\text{decay}} \Gamma\!_{0n} \frac{M_0}{m_X}
  \frac{q_n}{q_{0n}} \frac{B_{Ln}^2(q_n)}{B_{Ln}^2(q_{0n})}
\end{equation}
with $\Gamma\!_0 = \Gamma\!_\text{tot}(M_0)$. Here $q_n$ represents the
breakup momentum of the particular two-body decay and $q_{0n} =
q_n(M_0)$. If required by the data, for some waves an additional
coherent exponential background term of the form $e^{-\alpha q^2}$ was
added. The waves selected for the mass-dependent fit exhibit either
significant amplitudes or rapid phase motions in the 1.7\gevcc\ mass
region.

\subsection{Results}

\begin{figure}[!b]
  \centering
  \includegraphics[width=\columnwidth]{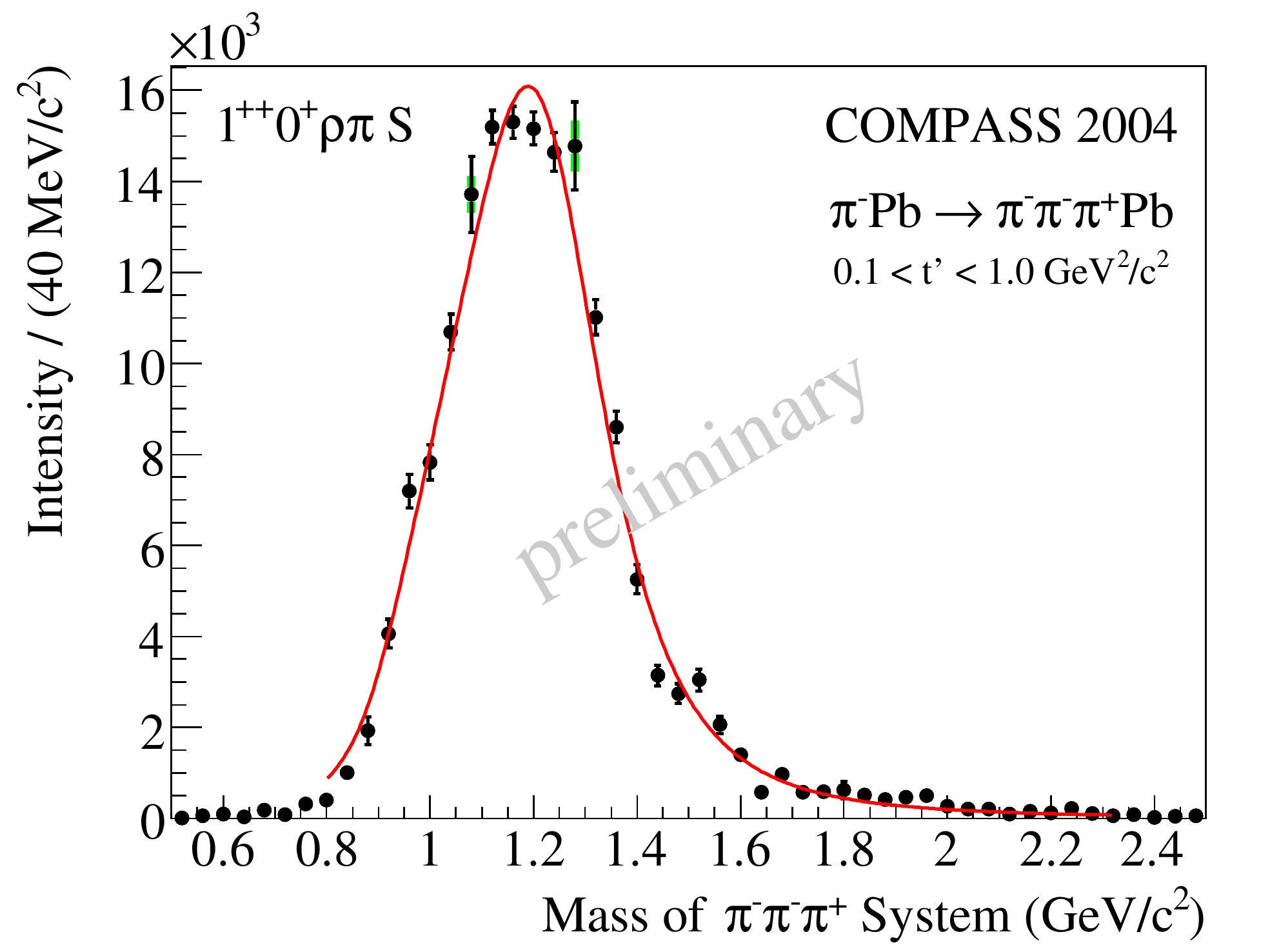}
  \caption{Intensity of the \wavespec{1}{+}{+}{0}{+}{\Pr\Ppi}{S}
    partial wave. The wave exhibits a broad peak from the
    $\Paone(1260)$. The continuous line shows the result of the
    mass-dependent fit.}
  \label{fig:result_a1}
\end{figure}

\Figref{result_a1} shows the most dominant wave,
\wavespec{1}{+}{+}{0}{+}{\Pr\Ppi}{S}. The data points with statistical
error bars are the result of the mass-independent fit. The wave
intensity exhibits a broad structure around 1.2\gevcc\ which is the
$\Paone(1260)$. The continuous line represents the result of the
mass-dependent $\chi^2$-fit of the intensities and phase differences
of six waves (see \tabref{waveset}). In this fit the peak is well
described using a Breit-Wigner parameterization according to
Bowler~\cite{bw_bowler} and a small exponential background which could
be caused by the Deck-effect~\cite{deck}.

\begin{figure}[!t]
  \centering
  \includegraphics[width=\columnwidth]{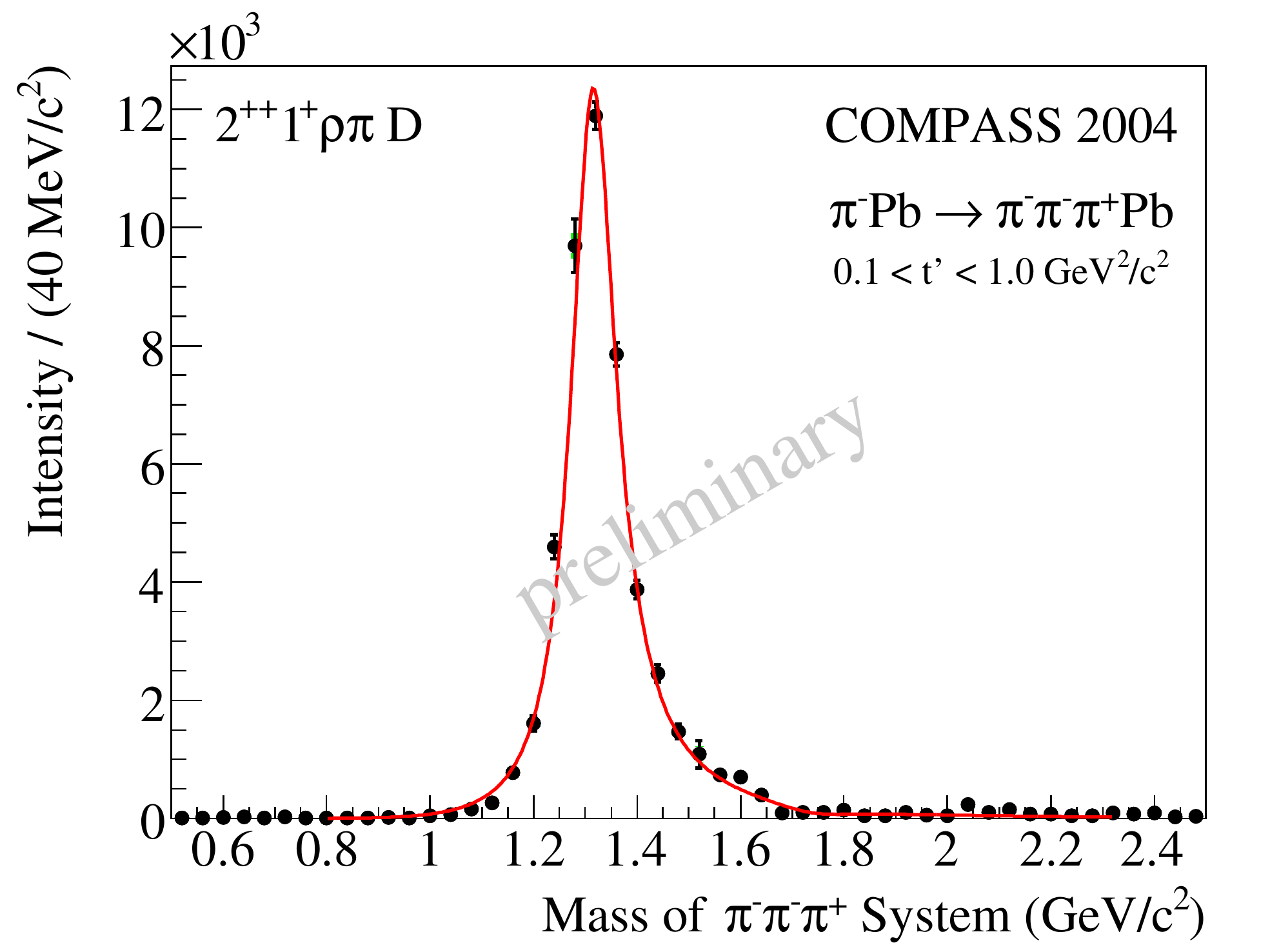}
  \caption{Intensity of the \wavespec{2}{+}{+}{1}{+}{\Pr\Ppi}{D}
    partial wave. The wave is dominated by a narrow $\Patwo(1320)$
    peak. The continuous line shows the result of the mass-dependent fit.}
  \label{fig:result_a2}
\end{figure}

The second most intense wave is the
\wavespec{2}{+}{+}{1}{+}{\Pr\Ppi}{D} wave which is depicted in
\figref{result_a2}. It is also the strongest $M = 1$ wave and can be
large only at high $t'$. The sharp peak of the $\Patwo(1320)$ is
described by a Breit-Wigner, where the mass-dependent total width
$\Gamma\!_\text{tot}(m_X)$ takes into account the two dominant decay
modes of the $\Patwo(1320)$ to $[\Pr\Ppi]D$ and $[\Peta\Ppi]D$ with
strength according to PDG~\cite{pdg}. In order to reproduce the
high-mass tail of the wave and its interference with the
\wavespec{1}{+}{+}{0}{+}{\Pr\Ppi}{S} and the
\wavespec{2}{-}{+}{0}{+}{\Pftwo\Ppi}{S} waves, a second Breit-Wigner
for the $\Patwo(1700)$ with its parameters fixed to the PDG
values~\cite{pdg} was added to the fit.

\begin{figure}[!t]
  \centering
  \includegraphics[width=\columnwidth]{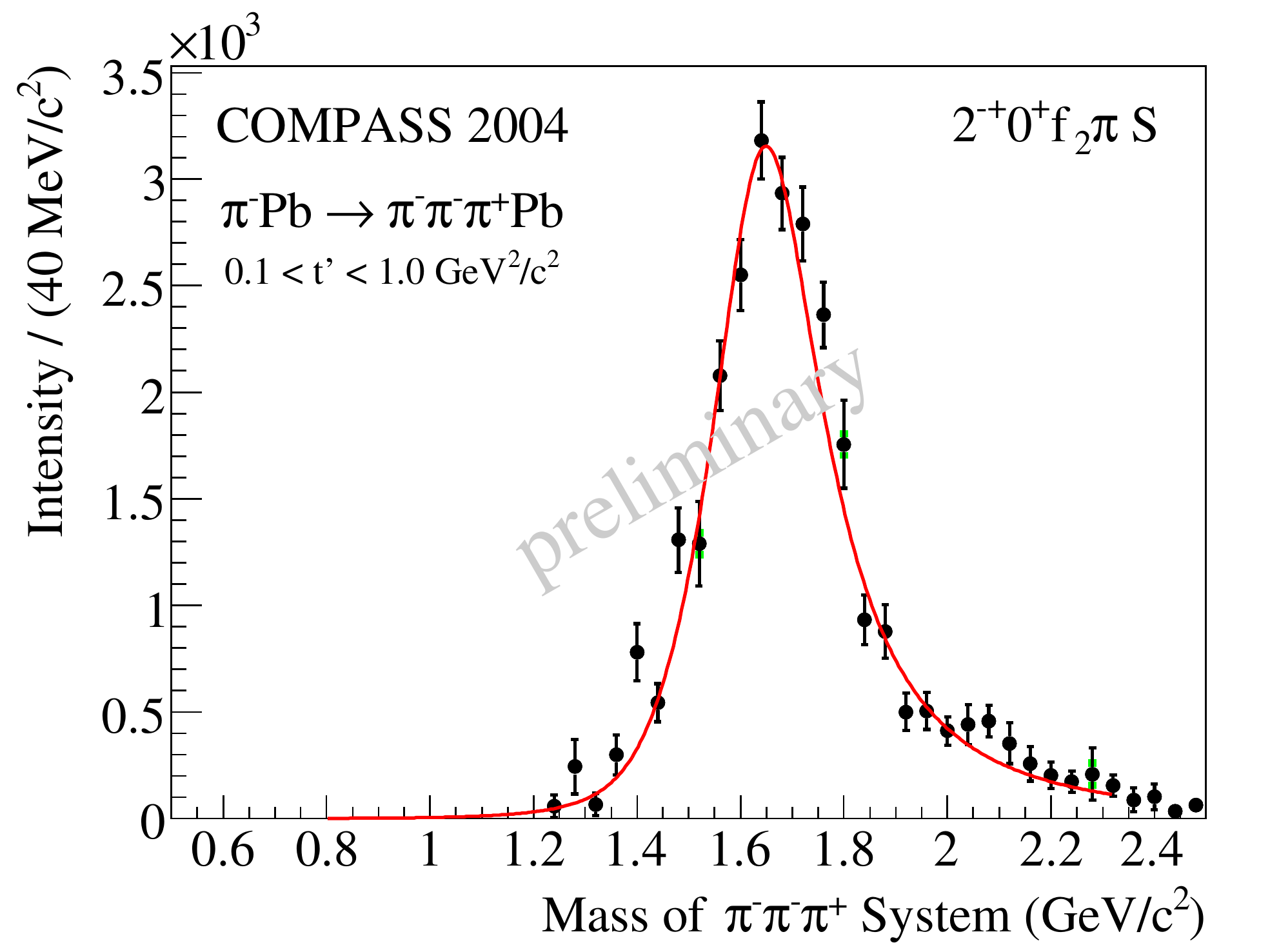}
  \caption{Intensity of the \wavespec{2}{-}{+}{0}{+}{\Pftwo\Ppi}{S}
    partial wave. The wave exhibits a broad $\Ppitwo(1670)$ peak. The
    continuous line represents the result of the mass-dependent fit.}
  \label{fig:result_pi2}
\end{figure}

The third significant wave is the
\wavespec{2}{-}{+}{0}{+}{\Pftwo\Ppi}{S} wave shown in
\figref{result_pi2}. A Breit-Wigner corresponding to the
$\Ppitwo(1670)$ fits the data well. Its total width was parameterized
assuming 60~\% $[\Pftwo\Ppi]S$ and 40~\% $[\Pr\Ppi]P$ decays,
neglecting contributions below 10~\%.  

\begin{figure}[!t]
  \centering
  \includegraphics[width=\columnwidth]{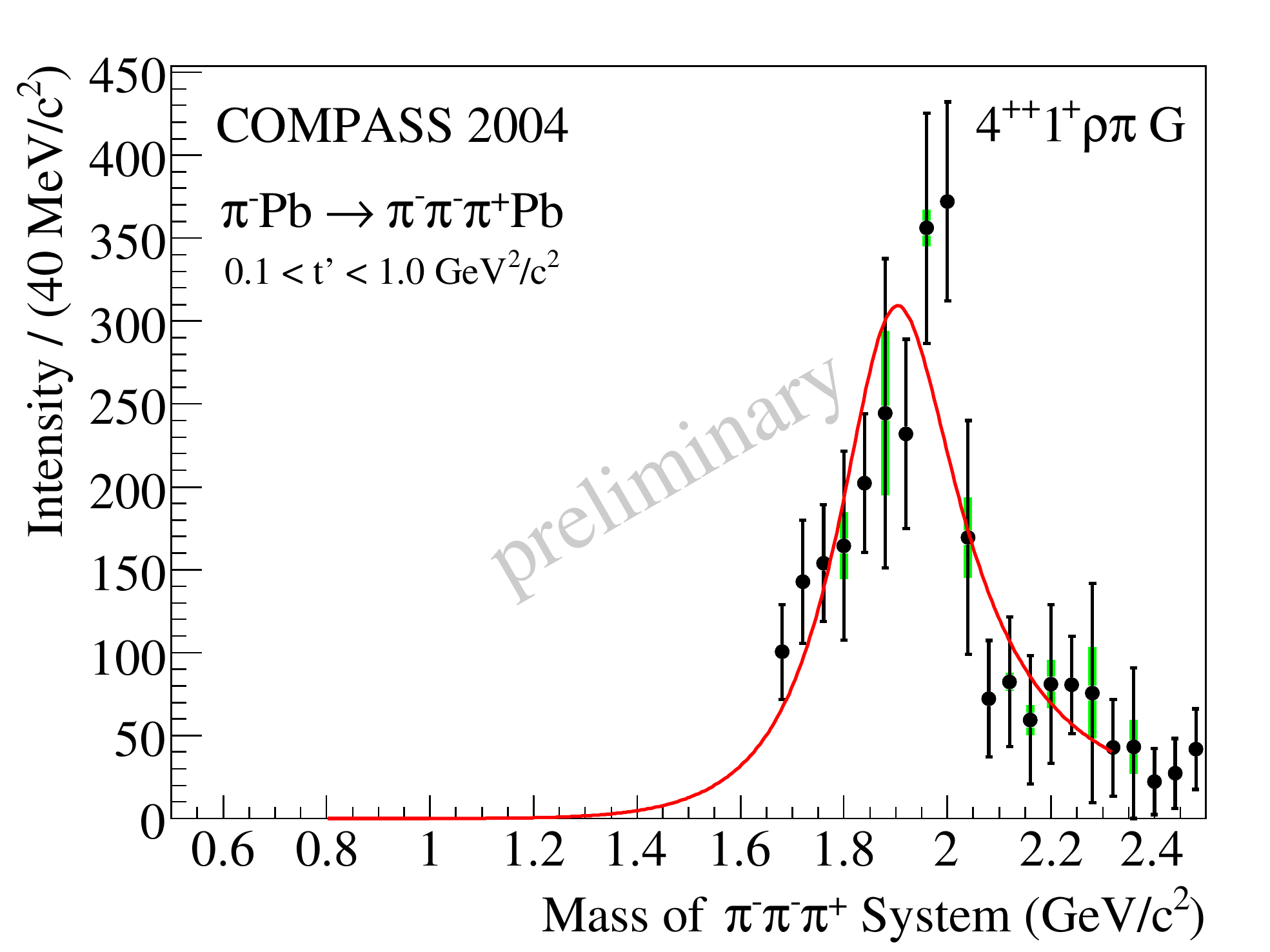}
  \caption{Intensity of the \wavespec{4}{+}{+}{1}{+}{\Pr\Ppi}{G}
    partial wave. The broad peak around 1.9\gevcc\ is attributed to
    the $\Pafour(2040)$. The continuous line shows the result of the
    mass-dependent fit.}
  \label{fig:result_a4}
\end{figure}

\begin{figure}[!t]
  \centering
  \includegraphics[width=\columnwidth]{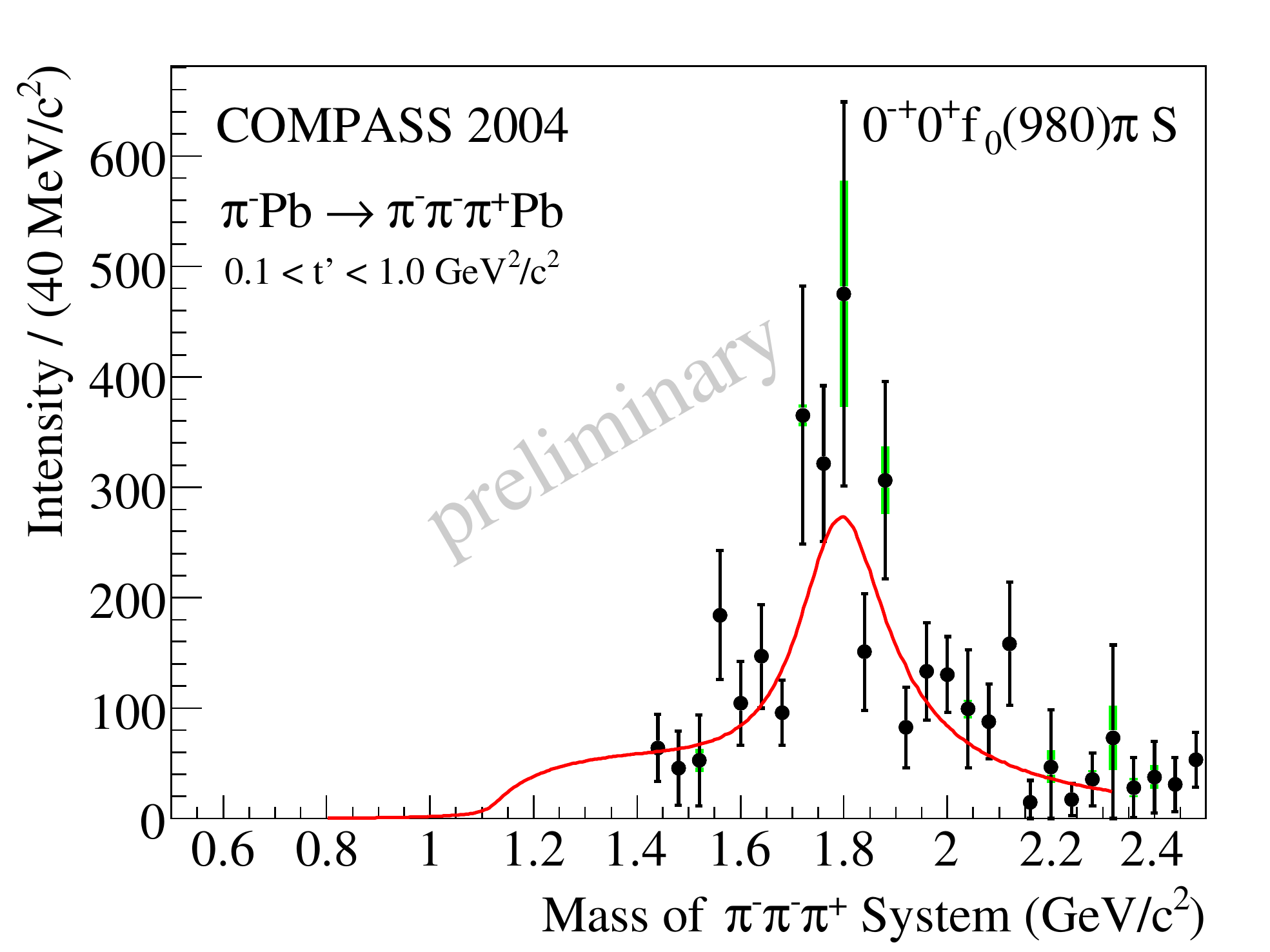}
  \caption{Intensity of the
    \wavespec{0}{-}{+}{0}{+}{\Pfzero(980)\Ppi}{S} partial wave. The
    structure around 1.8\gevcc\ is attributed to the $\Ppi(1800)$. The
    continuous line shows the result of the mass-dependent fit.}
  \label{fig:result_pi}
\end{figure}

In order to study the sensitivity of the PWA to small signals, two
low-intensity waves with known resonances were included in the
mass-dependent fit: the \wavespec{4}{+}{+}{1}{+}{\Pr\Ppi}{G} wave with
the $\Pafour(2040)$ (see \figref{result_a4}) and the
$0^{-+}0^+[\Pfzero(980) \linebreak \Ppi] S$ wave with the
$\Ppi(1800)$ (see \figref{result_pi}). The
\wavespec{4}{+}{+}{1}{+}{\Pr\Ppi}{G} wave peaks around 1.9\gevcc\ and
is fit by a Breit-Wigner amplitude using a constant width in its
denominator, because no branching fractions are known for the
$\Pafour(2040)$. The resulting $\Pafour$ mass of $1\,884 \pm
13(\text{stat.})^{+50}_{-2}(\text{syst.})$ is lower than the PDG
average~\cite{pdg}, but has a large systematic error. Even fainter in
intensity is the $\Ppi(1800)$ peak in the
\wavespec{0}{-}{+}{0}{+}{\Pfzero(980)\Ppi}{S} wave, but it exhibits a
clear phase motion with respect to the $\Ppitwo(1670)$ in the
\wavespec{2}{-}{+}{0}{+}{\Pftwo\Ppi}{S} wave (not shown here). A
Breit-Wigner with constant total width on top of a background
describes the data.

\begin{figure}[!t]
  \centering
  \includegraphics[width=\columnwidth]{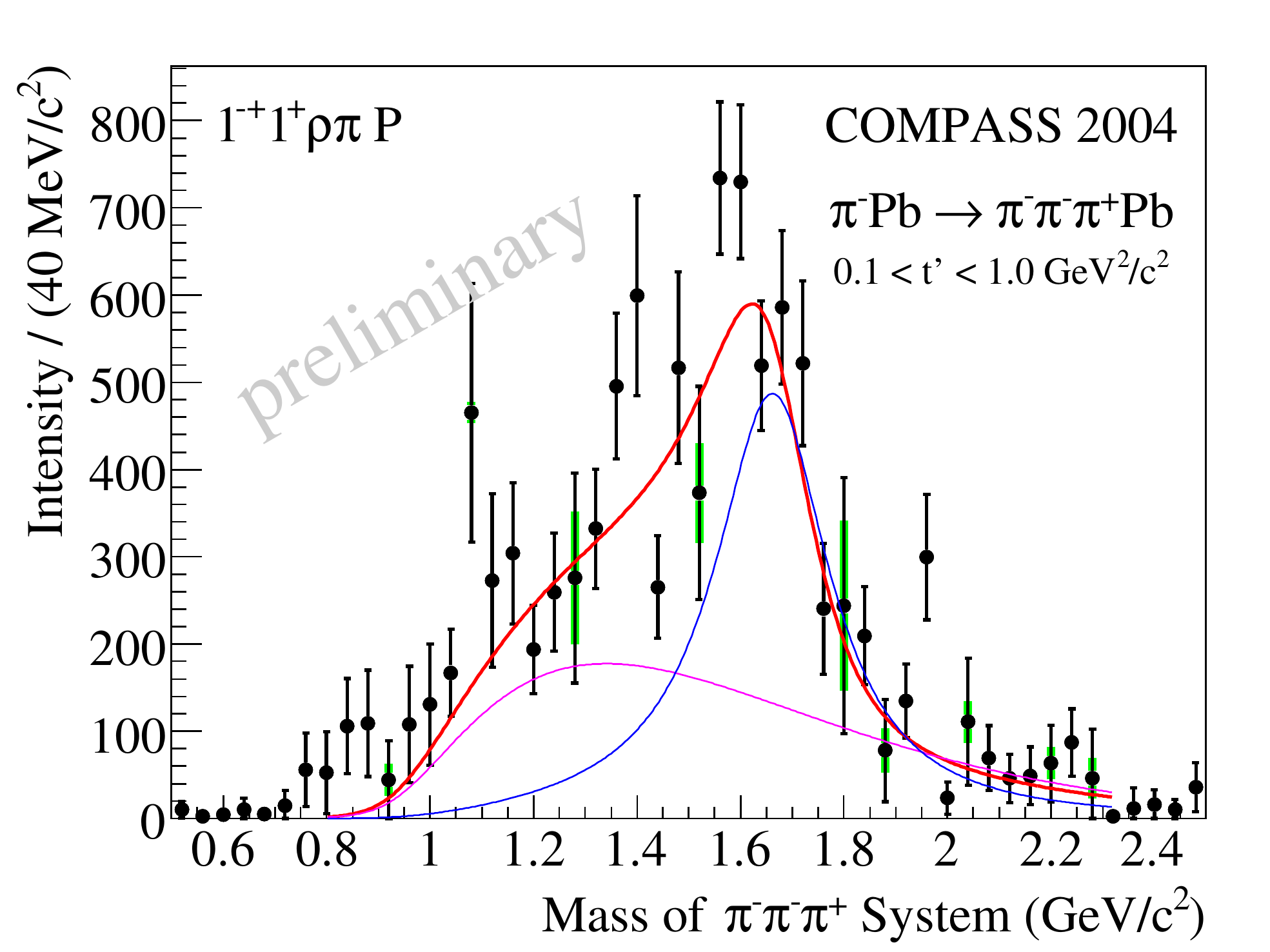}
  \caption{Intensity of the spin-exotic
    \wavespec{1}{-}{+}{1}{+}{\Pr\Ppi}{P} partial wave. The
    distribution exhibits a broad bump around 1.7\gevcc. The red curve
    shows the result of the mass-dependent fit with one Breit-Wigner
    for the $\Ppione(1600)$ (blue curve) on top of a background
    (purple curve).}
  \label{fig:result_pi1}
\end{figure}

\begin{figure}[!t]
  \centering
  \includegraphics[width=\columnwidth]{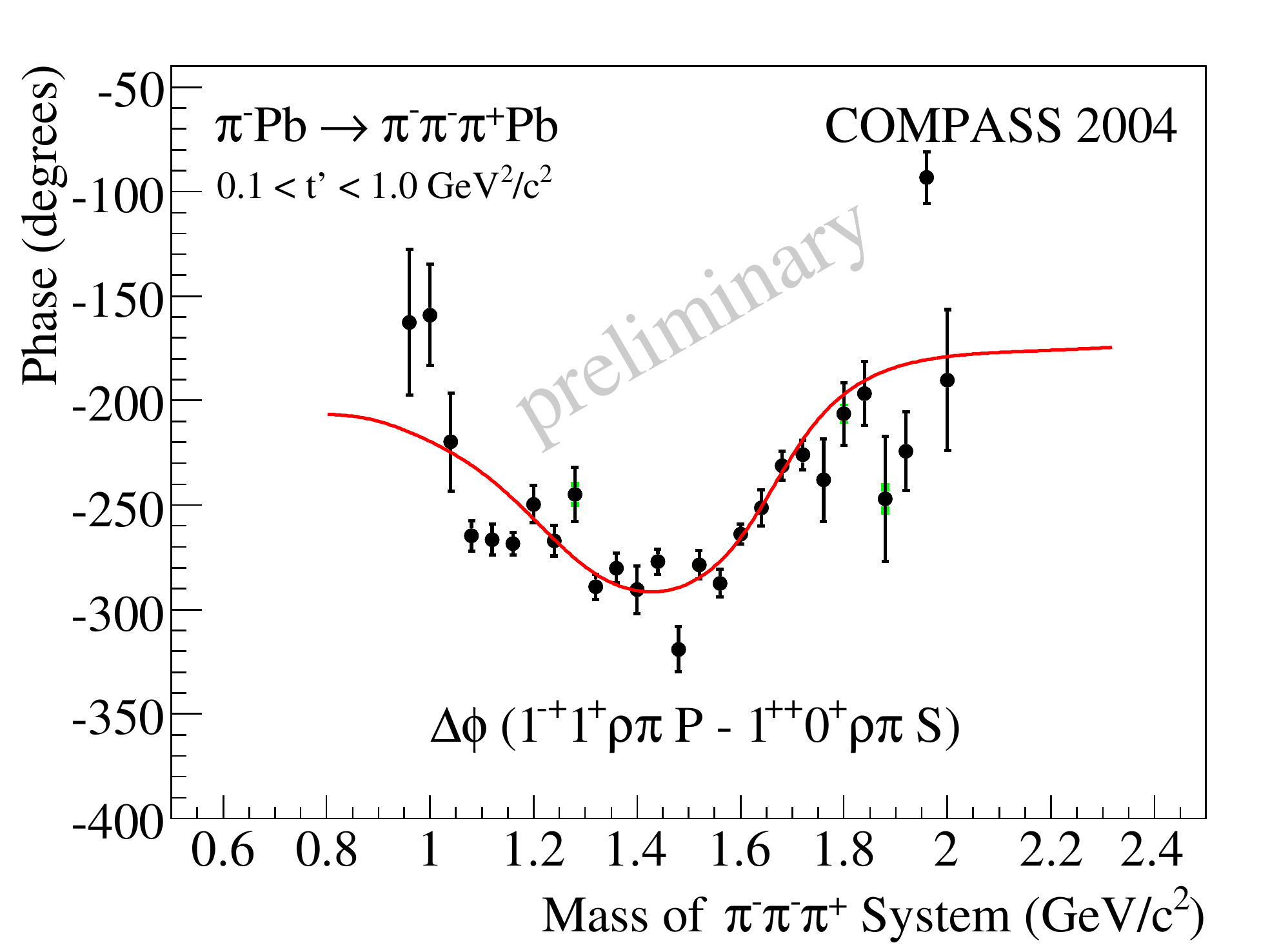}
  \caption{Phase difference of the
    \wavespec{1}{-}{+}{1}{+}{\Pr\Ppi}{P} and the
    \wavespec{1}{+}{+}{0}{+}{\Pr\Ppi}{S} partial waves. A clear phase
    motion is seen in the region around 1.7\gevcc\ which can be
    explained by a $\Ppione(1600)$ resonance interfering with the tail
    of the $\Paone(1260)$. The continuous line shows the result of the
    mass-dependent fit.}
  \label{fig:result_pi1-a1}
\end{figure}

\begin{figure}[!t]
  \centering
  \includegraphics[width=\columnwidth]{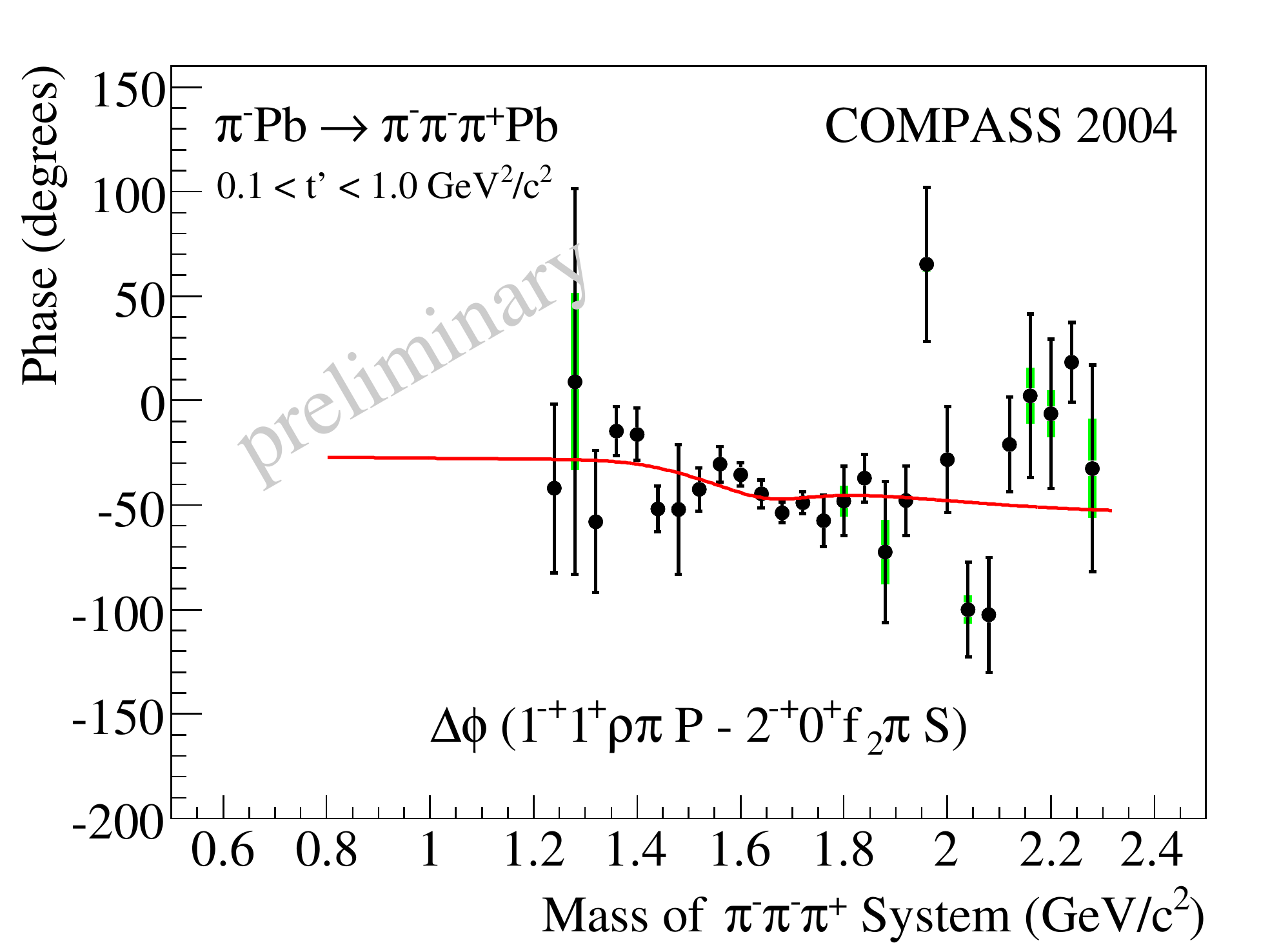}
  \caption{Phase difference of the
    \wavespec{1}{-}{+}{1}{+}{\Pr\Ppi}{P} and the
    \wavespec{2}{-}{+}{0}{+}{\Pftwo\Ppi}{S} partial waves. The phase
    difference is virtually flat which can be explained by a
    $\Ppione(1600)$ resonance with resonance parameters similar to the
    $\Ppitwo(1670)$. The continuous line shows the result of the
    mass-dependent fit.}
  \label{fig:result_pi1-pi2}
\end{figure}

\Figref{result_pi1} shows the intensity of the spin-exotic
\wavespec{1}{-}{+}{1}{+}{\Pr\Ppi}{P} wave. The distribution exhibits a
broad bump at 1.7\gevcc\ and a shoulder at lower masses. The phase
difference between this wave and the
\wavespec{1}{+}{+}{0}{+}{\Pr\Ppi}{S} wave, as depicted in
\figref{result_pi1-a1}, exhibits a rising phase motion around
1.7\gevcc, in the tail region of the $\Paone(1260)$, thus indicating a
resonant behavior of the spin-exotic wave. This is supported by the
flat phase difference with respect to the
\wavespec{2}{-}{+}{0}{+}{\Pftwo\Ppi}{S} wave as shown in
\figref{result_pi1-pi2} which can be explained by a $1^{-+}$ resonance
with mass and width similar to that of the $\Ppitwo(1760)$ which
dominates the $2^{-+}$ wave. The intensity and the phase differences
of the $1^{-+}$ amplitude are fit by a Breit-Wigner with constant
total width on top of an exponential background that describes the
low-mass shoulder. The extracted resonance parameters of $M_0 = 1\,660
\pm 10(\text{stat.})^{+0}_{-64}(\text{syst.})\mevcc$ and $\Gamma\!_0 =
269 \pm 21(\text{stat.})^{+42}_{-64}(\text{syst.})\mevcc$ are in
agreement with the disputed $\Ppione(1600)$.

\Tabref{fitResult} summarizes the resonance parameters extracted by
the mass-dependent fit which are in good agreement with the PDG
values~\cite{pdg}. The table also gives the intensities of the
resonant part of the particular waves integrated over the mass range
from 0.80 to 2.32\gevcc\ and normalized to the total intensity of the
waves in the mass-dependent fit, which corresponds to $38.7(2)~\%$ of
the total acceptance-corrected data sample in the mass range.

\begin{table*}
  \caption{Preliminary resonance parameters and intensities of the
    specified decay channels for the six waves included in the
    mass-dependent
    fit. The first uncertainty corresponds to the statistical error, the
    second one to the systematic error.}
  \label{tab:fitResult}
  \begin{center}
  \begin{tabular}{lllll}
    \toprule
    \textbf{State} &
    \textbf{Mass} &
    \textbf{Width} &
    \textbf{Intensity} &
    \textbf{Channel} \\
    &
    \textbf{$\boldmath{[\text{MeV}\! / c^2]}$} &
    \textbf{$\boldmath{[\text{MeV}\! / c^2]}$} &
    \textbf{$\boldmath{[\text{\%}]}$} &
    \textbf{$\boldmath{J^{PC}M^\epsilon[\text{decay}]L}$} \\
    \midrule[0.08em]
    $\Paone(1260)$  & $1\,255 \pm 6 ^{+7}_{-17}$  & $367 \pm 9 ^{+28}_{-25}$  & $67 \pm 3 ^{+4}_{-20}$        & \wavespec{1}{+}{+}{0}{+}{\Pr\Ppi}{S} \\[1.5ex]
    $\Patwo(1320)$  & $1\,321 \pm 1 ^{+0}_{-7}$   & $110 \pm 2 ^{+2}_{-15}$   & $19.2 \pm 0.6 ^{+0.3}_{-2.2}$ & \wavespec{2}{+}{+}{1}{+}{\Pr\Ppi}{D} \\[1.5ex]
    $\Ppitwo(1670)$ & $1\,658 \pm 3 ^{+24}_{-8}$  & $271 \pm 9 ^{+22}_{-24}$  & $10.0 \pm 0.4 ^{+0.7}_{-0.7}$ & \wavespec{2}{-}{+}{0}{+}{\Pftwo\Ppi}{S} \\[1.5ex]
    $\Pafour(2040)$ & $1\,885 \pm 13 ^{+50}_{-2}$ & $294 \pm 25 ^{+46}_{-19}$ & $1.0 \pm 0.3 ^{+0.1}_{-0.1}$  & \wavespec{4}{+}{+}{1}{+}{\Pr\Ppi}{G} \\[1.5ex]
    $\Ppi(1800)$    & $1\,785 \pm 9 ^{+12}_{-6}$  & $208 \pm 22 ^{+21}_{-37}$ & $0.8 \pm 0.1 ^{+0.3}_{-0.1}$  & \wavespec{0}{-}{+}{0}{+}{\Pfzero\Ppi}{F} \\[1.5ex]
    $\Ppione(1600)$ & $1\,660 \pm 10 ^{+0}_{-64}$ & $269 \pm 21 ^{+42}_{-64}$ & $1.7 \pm 0.2 ^{+0.9}_{-0.1}$  & \wavespec{1}{-}{+}{1}{+}{\Pr\Ppi}{P} \\
    \bottomrule
  \end{tabular}
  \end{center}
\end{table*}
 
Several studies were performed to test the stability of the $1^{-+}$
wave with respect to various assumptions made in the analysis, e.g. by
adding and removing waves, varying cuts or initial values of the fit
parameters, and shifting the mass binning in the mass-independent
fit. In addition the dependence of the fit on the rank $N_r$ was
studied. The fit with $N_r = 1$ shows an increase of the flat
background wave from 5.8~\% for $N_r = 2$ to 19~\% with respect to the
total acceptance-corrected data sample in the mass range from 0.5 to
2.5\gevcc. For $N_r = 3$ this value drops to 1.2~\%. This fit,
however, exhibits larger bin-to-bin fluctuations. In both cases the
$1^{-+}$ intensity was not significantly altered in the region between
1.5 and 1.8\gevcc.

In~\cite{e852_pione1600_rho_2} it was shown that an inhomogeneous
acceptance not completely taken into account by the Monte Carlo
simulation or an incomplete wave set may lead to leakage of intensity
of non-exotic waves into the $1^{-+}1^+$ wave. This effect was studied
by generating Monte Carlo events using the parameters of the 16 most
dominant waves excluding the $1^{-+}$ wave. The Monte Carlo data were
then processed by the same PWA used for the real data. The amount of
intensity leaking from the 16~major waves into the
\wavespec{1}{-}{+}{1}{+}{\Pr\Ppi}{P} wave was found to be below 5~\%
of the $1^{-+}$ intensity and is therefore negligible. This is mainly
due to the flat angular acceptance and the better resolution of
COMPASS, both well reproduced by the simulations.

In the mass-dependent fit the low mass shoulder in the
\wavespec{1}{-}{+}{1}{+}{\Pr\Ppi}{P} wave is accounted for by a
non-resonant exponential background, which is possibly caused by a
Deck-like effect~\cite{deck}. In an alternative attempt to describe
the shoulder an additional Breit-Wigner for the $\Ppione(1400)$ with
its parameters fixed to the PDG values~\cite{pdg} was included in the
mass-dependent fit. This did not significantly alter the intensity or
the phase difference of any of the waves in the mass-dependent fit,
but slightly shifted the $\Ppione(1600)$ towards smaller masses which
was included in the systematic error.

Also using different parameterizations for the \Ppi\Ppi\ $S$-wave and
the $\Pr(770)$ and replacing the Zemach tensors by rotation functions
with relativistic corrections~\cite{relat_dfunc} did not significantly
change the result. This is also true for tests, where the PWA was
performed without the $t'$-dependent factors in the partial-wave
amplitudes.

It cannot be ruled out, however, that the $\Ppione(1600)$ signal is an
artifact of the simplifying assumptions made in the PWA model. Final
state interactions in particular the Deck-mechanism were found to be
relevant in the \threepion\ system~\cite{deck_1,deck_2} and might
cause spurious signals.

\subsection{Run 2008/9 Data}

The analysis of the data from the 2004 pilot run marked only the
beginning of the hadron spectroscopy program at COMPASS. During the
years 2008 and 2009 COMPASS collected very large diffractive and
central production data sets using 190\gevc\ positive and negative
hadron beams mostly on liquid hydrogen targets. These data will make
it possible to study a number of channels with unprecedented precision.

\begin{figure}[!h]
  \centering
  \includegraphics[height=\columnwidth,angle=90]{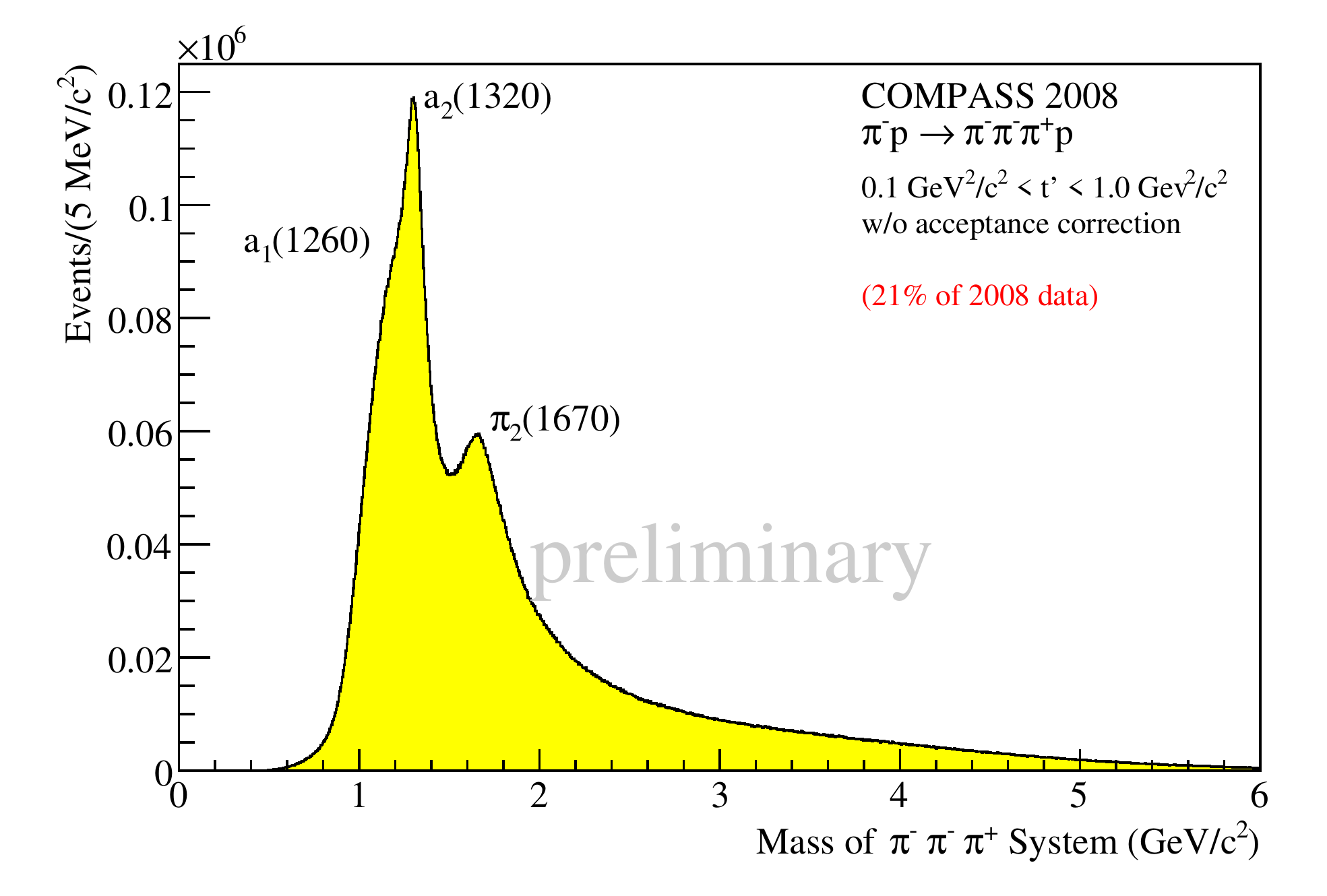}
  \caption{Invariant mass distribution \threepion\ final states
    diffractively produced by a \Ppim\ beam on a liquid hydrogen
    target; based on 21~\% of the 2008 data set.}
  \label{fig:diffr_3pi_2008}
\end{figure}

Several analyses have started: \Figref{diffr_3pi_2008} shows the
\linebreak \threepion\ invariant mass spectrum after event
selection. The tremendous boost in statistics will enable COMPASS to
extract precise resonance parameters, to study the $t'$-depen\-dence of
resonance production, and to hopefully settle the $\Ppione(1600)$
issue in the diffractive \threepion\ channel. This analysis will be
accompanied by a PWA of the \Ppim\Ppiz\Ppiz\ final state which will
provide important internal cross checks of our results. Furthermore
the PWA will be extended to the region of $t' < 0.1\gevcsq$. Data sets
with different targets will give access to possible nuclear effects in
diffractive production.

\begin{figure}[!h]
  \centering
  \includegraphics[width=\columnwidth]{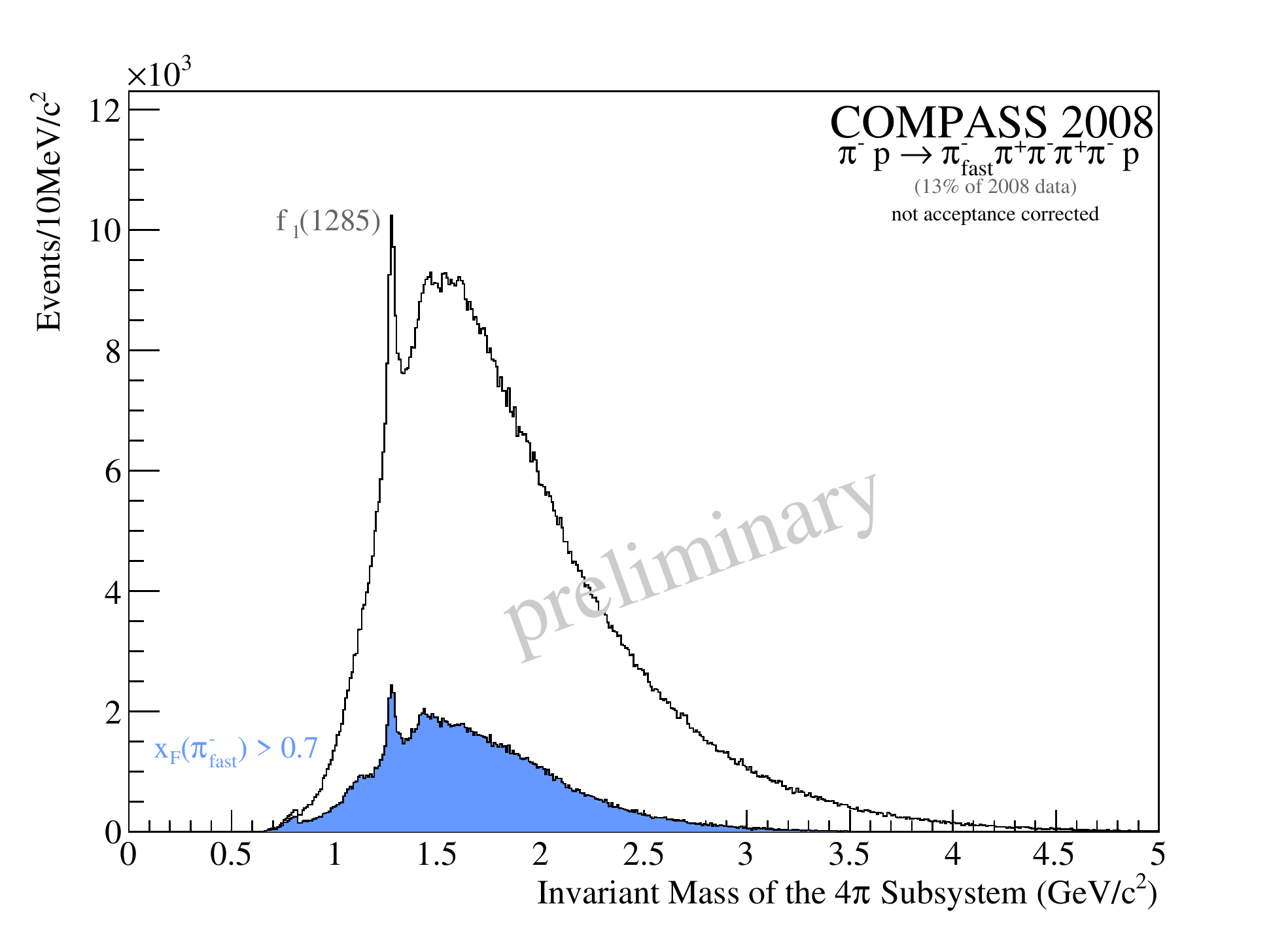}
  \caption{Invariant mass distribution of the \fourpion\ system
    produced in $\Ppim\Pp \to \Ppim_\text{fast} (\fourpion)
    \Pp_\text{recoil}$ with and without the cut
    $\xF(\Ppim_\text{fast}) > 0.7$; based on 13~\% of the 2008 data
    set.}
  \label{fig:cp_4pi_2008}
\end{figure}

In addition COMPASS will search for scalar and tensor glueballs in
centrally produced \fourpion\ systems. \Figref{cp_4pi_2008} depicts the
\fourpion\ invariant mass with and without selecting centrally
produced event via the cut $\xF(\Ppim_\text{fast}) > 0.7$. A clear
signal from the $\Pfone(1285)$ is seen.

The $\Pfone\Ppi$ and maybe even the $\Pbone\Ppi$ channel, into which
the $1^{-+}$ hybrid is predicted to preferentially decay, can also be
studied in charged five pion final states produced in $\Ppi A \to
\fivepion A$. Furthermore this channel gives access to higher masses
above 2\gevcc, where the PDG lists many states that need
confirmation~\cite{pdg}.

Finally COMPASS will search for possible hybrids and glueballs in
kaonic final states produced in reactions like $\Ppim\Pp \to
\Ppim\PKp\PKm\Pp$, $\Ppim\PKs\PKs\Pp$, or $\Ppim\PKs\PKl\Pp$. Moreover
the CEDAR detectors enable COMPASS to measure diffraction of kaon
beam, e.g. in $\PKm\Pp \to \PKm\Ppip\Ppim\Pp$.

\section{Summary}

COMPASS has started its hadron spectroscopy program with the main goal
to precisely explore the light-meson sector in order to search for
states with gluonic degrees of freedom like hybrids and glueballs and
to settle the properties of controversial states like the
$\Ppione(1400)$, $\Ppione(1600)$, and the $\Pfzero(1500)$. At COMPASS
two production mechanisms, diffractive dissociation and central
production, can be measured in parallel using \Ppi, \PK, or \Pp\
beams. The spectrometer provides high acceptance and excellent
resolution for final states containing both charged and neutral
particles.

In a short first pilot run in 2004 COMPASS acquired a diffractive data
set using a 190\gevc\ \Ppim\ beam scattered off a lead target. A
partial wave analysis of the \threepion\ final state in the region of
$t' \in [0.1, 1.0]\gevcsq$ shows several of the well-known resonances
expected in this channel. In addition we observe a spin-exotic $\jpc =
1^{-+}$ state with significant intensity in the $\Pr\Ppi$ decay
channel in natural parity exchange at a mass of 1.66\gevcc. The mass
dependence of its phase and its resonance parameters are consistent
with the disputed $\Ppione(1600)$.

Based on the experience of the pilot run the spectrometer was upgraded
and improved. In particular a detector to measure the target recoil
was added to the setup. During dedicated runs in 2008 and 2009 COMPASS
accumulated diffractive and central production data sets of
unprecedented statistics using 190\gevc\ hadron beams. The bulk of the
data were taken with a liquid hydrogen target, but also nickel,
tungsten, and lead targets were measured. First analyses of
diffractive and central production reactions have started.

\begin{acknowledgement}
  This work is supported by the the German Bundesministerium für
  Bildung und Forschung, the Maier-Leibnitz-Labor der LMU und TU
  München, the DFG Cluster of Excellence \emph{Origin and Structure
    of the Universe}, and CERN-RFBR grant 08-02-91009.
\end{acknowledgement}

\end{document}